\documentclass[12pt, reqno]{amsart}

\usepackage[margin=.9in, a4paper, foot=.3in]{geometry}

\usepackage[font=small,justification=centering]{caption}

\usepackage{cite}

\usepackage{graphicx, color}

\definecolor{darkblue}{rgb}{0,0,.5}
\definecolor{darkred}{rgb}{.5,0,0}
\definecolor{darkgreen}{rgb}{0,0.5,0}

\usepackage{hyperref}
\hypersetup {
    pdfstartview=FitH,
    colorlinks,
    citecolor=darkgreen,
    linkcolor=darkred,
    urlcolor=darkblue
}

\usepackage{bm, bbm}
\usepackage{amssymb}

\usepackage[noBBpl]{mathpazo}
\usepackage[mathcal]{euscript}

\allowdisplaybreaks

\numberwithin{equation}{section}

\usepackage{mathtools}

\newcommand {\bil}[2]{(\, #1 \, | \, #2 \,)}
\newcommand {\diag}{\mathrm{diag}}
\newcommand {\ldbr}{[\![}
\newcommand {\oo}{{\overline 1}}
\newcommand {\oz}{{\overline 0}}
\newcommand {\rdbr}{]\!]}

\newcommand{\calR}{\mathcal R}

\newcommand {\bbE}{\mathbb E}
\newcommand {\bbZ}{\mathbb Z}

\newcommand {\rme}{\mathrm e}

\newcommand {\bbC}{\mathbb C}

\newcommand {\gothg}{\mathfrak g}
\newcommand {\gothh}{\mathfrak h}
\newcommand {\gothk}{\mathfrak k}

\newcommand {\glmn}{\mathfrak{gl}_{M | N}}
\newcommand {\hgothh}{\widehat{\mathfrak h}}
\newcommand {\hlslmn}{\widehat{\mathcal L}(\mathfrak{sl}_{M | N})}
\newcommand {\lslmn}{\mathcal L(\mathfrak{sl}_{M | N})}
\newcommand {\slmn}{\mathfrak{sl}_{M | N}}
\newcommand {\tgothh}{\widetilde{\mathfrak h}}

\newcommand {\tlslmn}{\widetilde{\mathcal L}(\mathfrak{sl}_{M | N})}
\newcommand {\uqg}{\mathrm U_q(\mathfrak g)}
\newcommand {\uqglmn}{\mathrm U_q(\mathfrak{gl}_{M | N})}
\newcommand {\uqlslm}{\mathrm U_q(\mathcal{L}(\mathfrak{sl}_M))}
\newcommand {\uqlslmn}{\mathrm U_q(\mathcal{L}(\mathfrak{sl}_{M | N}))}
\newcommand {\uqlslto}{\mathrm U_q(\mathcal{L}(\mathfrak{sl}_{2 | 1}))}
\newcommand {\uqslm}{\mathrm U_q(\mathfrak{sl}_M)}
\newcommand {\uqslmn}{\mathrm U_q(\mathfrak{sl}_{M | N})}

\newcommand {\End}{\mathrm{End}}
\newcommand {\id}{\mathrm{id}}

\newcommand {\str}{\mathrm{str}}

\title[Khoroshkin--Tolstoy approach for quantum superalgebras]{Khoroshkin--Tolstoy approach for quantum superalgebras}

\author{A. V. Razumov}
\address{Institute for High Energy Physics, NRC ``Kurchatov Institute", 142281 Protvino, Mos\-cow region, Russia}
\email{Alexander.Razumov@ihep.ru}

\begin{document}

\addtolength {\jot}{3pt}

\begin{abstract}
The central object of the quantum algebraic approach to the study of quantum integrable models is the universal $R$-matrix, which is an element of the completed tensor product of two copies of a quantum algebra. Various integrability objects are constructed by choosing representations for the factors of this tensor product. There are two approaches to constructing explicit expressions for the universal $R$-matrix. One is based on the use of Lusztig automorphisms, and the other is based on the concepts of normal ordering and $q$-commutator. In the case of a quantum superalgebra, we cannot use the first approach, since we do not know an explicit expression for the Lusztig automorphisms. The second approach can be used, although it requires some modifications. In this article, we present the necessary modification of the method and use it to find an $R$-operator for a quantum integrable system related to the quantum superalgebra $\uqlslmn$.
\end{abstract}

\maketitle

\tableofcontents

\section{Introduction}

The study of quantum integrable models is based on the constructing the corresponding integrability objects and solving the functional relations they satisfy.\footnote{For the terminology used for integrability objects, we refer to the papers \cite{BooGoeKluNirRaz14a, Raz21, Raz21a}.} Here, the most productive method is the quantum algebraic approach.\footnote{Previously, what we call quantum algebra was usually called quantum group. In fact, this object is an associative algebra, which in a sense is a deformation of the universal enveloping algebra of a Lie algebra. Nowadays, the term quantum algebra is more commonly used, and we adhere to this terminology.} The central object of this approach is the universal $R$-matrix, which is an element of the completed tensor product of two copies of the quantum algebra. Various integrability objects are constructed by choosing representations for the factors of this tensor product. The consistent application of the quantum algebraic approach was initiated by Bazhanov, Lukyanov and Zamolodchikov \cite{BazLukZam96, BazLukZam97, BazLukZam99}. They considered the quantum version of the KdV theory. Subsequently, this method turned out to be efficient for studying other quantum integrable models.

The general notion of a quantum algebra $\uqg$ was proposed by Drinfeld and Jimbo \cite{Dri85, Jim85} for the case when $\gothg$ is a Kac--Moody algebra with a symmetrizable generalized Cartan matrix. In the report \cite{Dri87}, Drinfeld described the quantum double construction of the universal $R$-matrix and presented an explicit expression for the quantum algebra $\mathrm U_q({\mathfrak{sl}_2})$. Later, the corresponding expression was found for the quantum algebra $\uqslm$ with arbitrary $M$ \cite{Ros89}, and for a general quantum algebra associated with a finite dimensional simple Lie algebra \cite{KirRes90, LevSoi91}. The explicit expression for the case of a quantum algebra associated with the simplest affine Lie algebra $\widehat{\mathfrak{sl}_2}$ was derived in the paper \cite{LevSoiStu93}, and for the general affine Lie algebra in the papers by Damiani \cite{Dam98, Dam00}. Alternative approach was proposed by Khoroshkin and Tolstoy \cite{TolKho92}. In fact, the main difference between the papers \cite{Dam98, Dam00} and \cite{TolKho92} lies in the method of constructing the suitable Poincar\'e--Birkhoff--Witt basis of the quantum algebra. In the paper \cite{Dam98} the Lusztig automorphisms \cite{Lus93} were used, and the authors of \cite{TolKho92, KhoTol93} used to this end the concepts of normal ordering $q$-commutator.

By generalizing the defining relations of quantum algebra appropriately, one arrives at quantum algebras associated with Lie superalgebras \cite{Yam94, Yam99}. The quantum double construction can be directly generalized to the case of such quantum algebras \cite{GouZhaBra93}. In the paper \cite{HakSed94}, it was used to find an explicit form of the universal $R$-matrix for the quantum superalgebra $\uqslmn$. The Khoroshkin--Tolstoy approach for the quantum superalgebras associated with finite dimensional superalgebras is presented in the paper \cite{KhoTol91}, and for the quantum superalgebras associated with Kac--Moody superalgebras with symmetrizable generalized Cartan matrix, in the papers \cite{KhoTol93a, KhoTol94, KhoTol94a}.

Explicit expressions for the universal $R$-matrix were used to construct the corresponding integrability objects. Here, $R$-operators \cite{KhoTol92, LevSoiStu93, ZhaGou94, BraGouZhaDel94, BraGouZha95, BooGoeKluNirRaz10, BooGoeKluNirRaz11}, monodromy operators, and $L$-operators were constructed \cite{BazTsu08,  BooGoeKluNirRaz10, BooGoeKluNirRaz11, BooGoeKluNirRaz13, Raz13, BooGoeKluNirRaz14a}. The corresponding families of functional relations were found \cite{BazHibKho02, Koj08, BazTsu08, BooGoeKluNirRaz14a, BooGoeKluNirRaz14b, NirRaz16a, Raz21, Raz21a}. Among other applications of the quantum algebraic approach, we would like to mention the description of the hidden fermionic structure of the XXZ model \cite{BooJimMiwSmiTak07, BooJimMiwSmiTak09, BooJimMiwSmi10}, and the derivation of equations satisfied by the reduced density operator of the quantum chain related to the quantum algebra $\uqslm$ \cite{KluNirRaz20, Raz20}.

The above list of applications of the quantum algebraic approach is by no means exhaustive. It should be noted that, apart from the paper \cite{BazTsu08}, it does not contain applications of this approach to quantum integrable systems related to quantum superalgebras. We have not found in the literature examples of derivation of an integrability object for a quantum integrable system related to a quantum superalgebra in the framework of the quantum algebraic approach, see, however, the paper \cite{IpZei14}, where some hints of the quantum algebraic approach were used for a special case of a quantum superalgebra. Note that the method based on usage of the Lusztig automorphisms is not applicable here, since we do not know their explicit form. Some necessary modifications of the Khoroshkin--Tolstoy approach are presented in the papers \cite{KhoTol93a, KhoTol94, KhoTol94a}, but, unfortunately, not all of them. In this paper, using the Khoroshkin--Tolstoy approach, we derive the explicit form of the $R$-operator for the evaluation vector representation of the quantum superalgebra $\uqlslmn$. As expected, up to normalization and arbitrariness in the grading, we obtain the Perk--Schultz $R$-operator \cite{PerSch81}. In fact, the main goal of this paper is to investigate the modification of the Khoroshkin--Tolstoy approach which is necessary for its use in constructing integrability objects for systems related to quantum superalgebras.

In section \ref{s:2} we give general information on the Lie superalgebras $\glmn$ and $\slmn$, describe their root structure, and present the symmetric bilinear form for the superalgebra $\slmn$. The quantum superalgebra $\uqglmn$ and its vector representation are defined in section \ref{s:3}. In the beginning of section \ref{s:4} we discuss the affinization of the Lie superalgebra $\slmn$, and then define the corresponding quantum superalgebra. The Khoroshkin--Tolstoy approach to the construction of Cartan--Weyl generators of this superalgebra is considered in section \ref{s:5}. The expression for the universal $R$-matrix is given in section~\ref{s:6}. Finally, the evaluation module corresponding to the vector representation of $\uqglmn$ is defined in section \ref{s:7}, and the explicit form of the corresponding $R$-operator is found in section \ref{s:8}.

We fix the deformation parameter $\hbar \in \bbC$ in such a way that $q = \exp(\hbar)$ is not a root of unity and assume that
\begin{equation*}
q^\nu = \exp (\hbar \nu)
\end{equation*}
for any $\nu \in \bbC$. We define $q$-numbers by the equation
\begin{equation*}
[\nu]_q = \frac{q^\nu - q^{- \nu}}{q - q^{-1}}, \qquad \nu \in \bbC.
\end{equation*}
Tensor products of quantum algebras are considered to be suitably completed, see the book \cite{ChaPre94}.

\section{\texorpdfstring{Lie superalgebras $\glmn$ and $\slmn$}{Lie superalgebras glM|N and slM|N}} \label{s:2}

We fix two positive integers $M$ and $N$ such that $M, N \ge 1$ and $M \ne N$, and denote by $\bbC_{M | N}$ the superspace\footnote{See appendix \ref{a:1} for relevant definitions and notation.} formed by $(M + N)$-tuples of complex numbers with the following grading. An element of $\bbC_{M | N}$ is even if its last $N$ components are zero, and odd if its first $M$ components are zero. For simplicity, we denote the Lie superalgebra $\mathfrak{gl}(\bbC_{M | N})$ as $\glmn$. We denote by $v_i$, $i = 1, \ldots, M + N$, the elements of the standard basis of $\bbC_{M | N}$. By definition,
\begin{equation*}
[v_i] = \oz, \quad i = 1, \ldots, M, \qquad [v_i] = \oo, \quad i = M + 1, \ldots, N.
\end{equation*}
It is convenient to use the notation
\begin{equation*}
[i] = [v_i], \qquad i = 1, \ldots, M + N.
\end{equation*}
The elements $\bbE_{i j} \in \glmn$, $i, j = 1, \ldots, M + N$, defined by the equation
\begin{equation*}
\bbE_{i j} v_k = v_i \delta_{j k},
\end{equation*}
form a basis of the Lie superalgebra $\glmn$. It is clear that the matrices of $\bbE_{i j}$ with respect to the standard basis of $\bbC_{M | N}$ are the usual matrix units, and we have
\begin{equation*}
\bbE_{i j} \bbE_{k l} = \delta_{j k} \bbE_{i l}.
\end{equation*}
It is also evident that
\begin{equation*}
[\bbE_{i j}] = [i] + [j].
\end{equation*}

As the Cartan subalgebra $\gothk$ of the Lie superalgebra $\glmn$ we take the subalgebra span\-ned by the elements $K_i = \bbE_{i i}$, $i = 1, \ldots, M + N$, which form its basis. Denote by $(\Xi_i)_{i = 1, \ldots, M + N}$ the dual basis of the space $\gothk^*$. For $X = \sum_{i = 1}^{M + N} c_i K_i \in \gothk$ we have
\begin{equation*}
[X, \, \bbE_{i j}] = (c_i - c_j) \, \bbE_{i j} = \langle\Xi_i -\Xi_j, \, X \rangle \, \bbE_{i j}.
\end{equation*}
Hence, $\bbE_{i j}$, $i \ne j$, is a root vector corresponding to the root $\Xi_i -\Xi_j$ and the root system of $\glmn$ is the set
\begin{equation*}
\Delta = \{\Xi_i -\Xi_j \mid i, j = 1, \ldots, M + N, \ i \ne j\}.
\end{equation*}
We choose as the system of simple roots the set
\begin{equation*}
\Pi = \{\Xi_i -\Xi_{i + 1} \mid i = 1, \ldots, M + N - 1\},
\end{equation*}
then the system of positive roots corresponding to $\Pi$ is
\begin{equation*}
\Delta_+ = \{\alpha_{i j} = \Xi_i -\Xi_j \mid 1 \le i < j \le M + N\}.
\end{equation*}
Certainly, the corresponding system of negative roots is $\Delta_- = -\Delta_+$. Denoting
\begin{equation*}
\alpha_i =  \alpha_{i, \, i + 1} =\Xi_i -\Xi_{i + 1}, \qquad i = 1, \ldots, M + N - 1,
\end{equation*}
we obtain
\begin{equation*}
\alpha_{i j} =  \sum_{k = 1}^{j - 1} \alpha_k, \qquad 1 \le i < j \le M + N.
\end{equation*}
Note that the root $\alpha_M$ is odd, all other simple roots are even, see for the definition the paper \cite{Kac77}. Denoting the parity of a root $\gamma \in \Delta$ by $[\gamma]$, we see that
\begin{equation*}
[\alpha_{i j}] = [\bbE_{i j}] = [i] + [j].
\end{equation*}
Like any root system, $\Delta$ defines a strict partial order $\prec$ on $\gothh^*$. Here, for $\alpha, \beta \in \gothh^*$ we have $\beta \prec \alpha$ if and only if $\alpha - \beta$ is the sum of positive roots.

Identifying an element of the Lie superalgebra $\glmn$ with its matrix with respect to the standard basis of the superspace $\bbC_{M | N}$, we define the supertrace of an element $X = (X_{i j})_{i, j = 1}^{M + N}$ by the relation
\begin{equation*}
\str \, X =  \sum_{i = 1}^{M + N} (-1)^{[i]} X_{i i}.
\end{equation*}
The Lie superalgebra $\slmn$ is a Lie subsuperalgebra of the Lie superalgebra $\glmn$ formed by the elements with zero supertrace. As the Cartan subalgebra $\gothh$ of the Lie superalgebra $\slmn$ we take the Lie subsuperalgebra $\gothh$ of the Lie superalgebra $\gothk$, which consists of the elements with zero supertrace. The elements  $h_i \in \gothh$, $i = 1, \ldots, M + N - 1$, defined by the equation
\begin{equation*}
h_i =  K_i - (-1)^{[i] + [i + 1]} K_{i + 1}
\end{equation*}
form a basis of $\gothh$. As the positive and negative roots we take the restriction of the positive and negative roots of the Lie superalgebra $\glmn$ to $\gothh$. The numbers
\begin{equation*}
a_{i j} =  \langle \alpha_j, \, h_i \rangle, \qquad i, j = 1 \ldots M + N -1,
\end{equation*}
are the entries of the Cartan matrix $A = (a_{i j})_{i, j = 1}^{M + N - 1}$ of the Lie superalgebra $\slmn$. The explicit form of the Cartan matrix $A$ is shown in figure \ref{f:cmslmn}.
\begin{figure}
\begin{equation*}
A = \left[ \begin{array}{rrrrr|r|rrrrr}
2 & -1 & & & & \\
-1 & 2 & -1 & & &\\
& \ddots & \ddots & \ddots & & \\
& & \ddots & \ddots & \ddots & \\
& & & -1 & 2 & -1 \\
\hline
& & & & -1 & 0 & 1 \\
\hline
& & & & & -1 & 2 & -1 \\
& & & & & & \ddots & \ddots & \ddots \\
& & & & & & & \ddots & \ddots & \ddots \\
& & & & & & & & -1 & 2 & -1 \\
& & & & & & & & & -1 & 2 \\
\end{array} \right]
\end{equation*}
\caption{The Cartan matrix of the Lie superalgebra $\slmn$ with distinguished $M$th row and $M$th column.}
\label{f:cmslmn}
\end{figure}
Putting
\begin{equation*}
d_i =  (-1)^{[i]}, \qquad i = 1, \ldots, M + N,
\end{equation*}
and multiplying the matrix $A$ from the left by the diagonal matrix
\begin{equation*}
D = \diag (d_1, \ldots, d_M, d_{M + 1}, \ldots, d_{M + N - 1}) = \diag (\underbracket[.5pt]{1, \, \ldots, 1}_M, \, \underbracket[.5pt]{-1, \, \ldots, \, -1}_{N - 1}),
\end{equation*}
we get the symmetrizied Cartan matrix $B = D A = (b_{i j})_{i, j = 1}^{M + N -1}$. For the explicit form of the matrix $B$ see figure \ref{f:scmslmn}.
\begin{figure}
\begin{equation*}
B = \left[ \begin{array}{rrrrr|r|rrrrr}
2 & -1 & & & & \\
-1 & 2 & -1 & & &\\
& \ddots & \ddots & \ddots & & \\
& & \ddots & \ddots & \ddots & \\
& & & -1 & 2 & -1 \\
\hline
& & & & -1 & 0 & 1 \\
\hline
& & & & & 1 & -2 & 1 \\
& & & & & & \ddots & \ddots & \ddots \\
& & & & & & & \ddots & \ddots & \ddots \\
& & & & & & & & 1 & -2 & 1 \\
& & & & & & & & & 1 & -2 \\
\end{array} \right]
\end{equation*}
\caption{The symmetrized Cartan matrix of the Lie superalgebra $\slmn$ with distinguished $M$th row and $M$th column.}
\label{f:scmslmn}
\end{figure}

Define a symmetric bilinear form $\bil {\cdot}{\cdot}$ on $\gothh$ by the equation
\begin{equation*}
\bil {h_i}{h_j} = a^{}_{i j} \, d^{-1}_j.
\end{equation*}
This form is nondegenerate and induces a symmetric bilinear form on $\gothh^*$. One can demonstrate that
\begin{equation*}
\bil {\alpha_i}{\alpha_j} = d_i \, a_{i j} = b_{i j}.
\end{equation*}
Below we often use the relations
\begin{gather*}
\bil {\alpha_{i j}}{\alpha_{j l}} = -d_j, \\*
\bil {\alpha_{i j}}{\alpha_{i l}} = d_i, \quad j \ne l, \qquad \bil {\alpha_{i j}}{\alpha_{i j}} = d_i + d_j, \qquad \bil {\alpha_{i j}}{\alpha_{k j}} = d_j, \quad i \ne k.
\end{gather*}
In fact, these are all nonzero cases.

\section{\texorpdfstring{Quantum superalgebra $\uqglmn$}{Quantum superalgebra Uq(glm|n)}} \label{s:3}

We define the quantum superalgebra $\uqglmn$ as a unital associative $\bbC$-superalgebra generated by the elements\footnote{We use capital letters to distinguish between generators of the quantum superalgebra $\uqglmn$ and the quantum superalgebra $\uqlslmn$.}
\begin{equation*}
E_i, \quad F_i, \quad i = 1, \ldots, M + N - 1, \qquad q^X, \quad X \in \gothk,
\end{equation*}
which obey the corresponding defining relations. The $\bbZ_2$-grading of the superalgebra $\uqglmn$ is defined on generators as
\begin{equation*}
[q^X] = \overline 0, \qquad [E_i] = [F_i] = \left\{ \begin{array}{ll}
\overline 0, & i \ne M, \\[.5em]
\overline 1, & i = M.
\end{array} \right.
\end{equation*}

Before giving the explicit form of the defining relations, introduce the notion of the $q$-supercommutator. The abelian group
\begin{equation*}
Q = \bigoplus_{i = 1}^{M + N - 1} \bbZ \, \alpha_i.
\end{equation*}
is called the root lattice of the Lie superalgebra $\glmn$. Assuming that
\begin{equation*}
q^X \in \uqglmn_0, \qquad E_i \in \uqglmn_{\alpha_i}, \qquad F_i \in \uqglmn_{-\alpha_i},
\end{equation*}
we endow $\uqglmn$ with a $Q$-grading. Now, for any elements $X \in \uqglmn_\alpha$ and $Y \in \uqglmn_\beta$ define the $q$-supercommutator by the equation
\begin{align*}
& \ldbr X, \, Y \rdbr =  X Y - (-1)^{[X] [Y]} q^{- \bil \alpha \beta} Y X = X Y - (-1)^{[\alpha] [\beta]} q^{- \bil \alpha \beta} Y X 
\intertext{if $\alpha, \beta \succ 0$, by the equation}
& \ldbr X, \, Y \rdbr =  X Y - (-1)^{[X] [Y]} q^{\bil \alpha \beta} Y X = X Y - (-1)^{[\alpha] [\beta]} q^{\bil \alpha \beta} X Y 
\intertext{if $\alpha, \beta \prec 0$, and by the equation}
& \ldbr X, \, Y \rdbr =  X Y - (-1)^{[X] [Y]} Y X = X Y - (-1)^{[\alpha] [\beta]} Y X
\end{align*}
if $\alpha \prec 0$ and $\beta \succ 0$, or $\alpha \succ 0$ and $\beta \prec 0$.

The defining relations of the quantum superalgebra $\uqglmn$ have the form
\begin{gather*}
q^0 = 1, \qquad q^{X_1} q^{X_2} = q^{X_1 + X_2}, \\
q^X E_i q^{-X} = q^{\langle \alpha_i, \, X \rangle} E_i, \qquad q^X F_i q^{-X} = q^{-\langle \alpha_i, \, X \rangle} F_i, \\
\ldbr E_i, \, F_j \rdbr = \delta_{i j} \frac{q_i^{H_i} - q_i^{- H_i}}{q_i^{} - q_i^{- 1}}, \\
\ldbr E_i, \, E_j \rdbr = 0, \qquad  \ldbr F_j, \, F_i \rdbr = 0, \qquad \mbox{ if } \bil {\alpha_j} {\alpha_i} = 0,  \\
\ldbr E_i, \, \ldbr E_i, \, E_{i - 1} \rdbr \rdbr = 0, \qquad \ldbr \ldbr F_{i - 1}, \, F_i \rdbr, \, F_i \rdbr = 0, \qquad \mbox{ if } i \ne 1 \mbox{ and } \bil {\alpha_i} {\alpha_i} \ne 0, \hspace{2.8em} \\
\ldbr E_i, \, \ldbr E_i, \, E_{i + 1} \rdbr \rdbr = 0, \qquad \ldbr \ldbr F_{i + 1}, \, F_i \rdbr, \, F_i \rdbr = 0, \qquad \mbox{ if } i \ne M + N \mbox{ and } \bil {\alpha_i} {\alpha_j} \ne 0, \\
\ldbr \ldbr \ldbr E_{M - 1}, \, E_M \rdbr, \, E_{M + 1} \rdbr, \, E_M \rdbr = 0, \qquad \ldbr F_M, \, \ldbr F_{M + 1}, \, \ldbr F_M, \, F_{M - 1} \rdbr \rdbr \rdbr = 0, 
\end{gather*}
where $i, j = 1, \ldots, M + N - 1$. Here and below we use the notation
\begin{equation*}
q_i =  q^{d_i}.
\end{equation*}
It is useful to have in mind that
\begin{equation*}
q_i^{} + q_i^{-1} = q + q^{-1}
\end{equation*}
and
\begin{equation*}
q^{}_i - q^{-1}_i = d_i (q - q^{-1}).
\end{equation*}
A general description of the defining relations of quantum superalgebras is given in the paper~\cite{Yam99}.

An element $a \in \uqglmn$ is called a root vector corresponding to a root $\gamma$ of the Lie superalgebra $\glmn$ if $a \in \uqglmn_\gamma$. In particular, $E_i$ and $F_i$ are root vectors corresponding to the roots $\alpha_i$ and $- \alpha_i$, respectively. It is possible to  construct linearly independent root vectors corresponding to all roots of the Lie superalgebra $\glmn$. To this end, following by Jimbo \cite{Jim86a}, we introduce the elements $E_{ij}$ and $F_{ij}$, $1 \le i < j \le M + N$, with the help of the relations
\begin{gather*}
E_{i, \, i + 1} = E_i, \qquad F_{i, \, i + 1} = F_i, \\
E_{i, \, k + 1} = \ldbr  E_{i k}, \, E_{k, \, k + 1} \rdbr, \qquad F_{i, \, k + 1} = \ldbr F_{k, \, k + 1}, \, F_{i k} \rdbr, \qquad k > i.
\end{gather*} 
In a more explicit form, the last two equations look as
\begin{equation*}
E_{i, \, k + 1} = E_{i k} \, E_{k, \, k + 1} - q_k \, E_{k, \, k + 1} \, E_{i k}, \qquad F_{i, k + 1 } = F_{k, \, k + 1} \, F_{i k} - q_k^{-1} F_{i k} \, F_{k, \, k + 1}.
\end{equation*}
Note that we have
\begin{equation*}
[E_{i j}] = [i] + [j],
\end{equation*}
in particular,
\begin{equation*}
[E_i] = [F_i] = [i] + [i + 1].
\end{equation*}
It is clear that $E_{i j}$ and $F_{i j}$ are root vectors corresponding to the roots $\alpha_{i j}$ and $- \alpha_{i j}$, respectively. These vectors are linearly independent, and together with the elements $q^X$, $X \in \gothk$, are called the Cartan--Weyl generators of the quantum superalgebra $\uqglmn$. It appears that the ordered monomials constructed from the Cartan--Weyl generators form a Poincar\'e--Birkhoff--Witt basis of the quantum superalgebra $\uqglmn$.

In the present paper we will work with the so called vector representation $\pi$ of the quantum superalgebra $\uqglmn$ on the superspace $\bbC_{M | N}$ defined by the relations
\begin{align}
& \pi(q^{\nu K_i}) = q^\nu \bbE_{i i} + \sum_{\substack{k = 1 \\ k \ne i}}^{M + N} \bbE_{k k}, \qquad i = 1, \ldots, M + N, \label{pik} \\
& \pi(E_i) = \bbE_{i, \, i + 1}, \qquad \pi(F_i) = \bbE_{i + 1, \, i}, \qquad i = 1, \ldots, M + N - 1. \label{piepif}
\end{align}
This representation will be used in section \ref{s:7} to construct the evaluation vector representation of the quantum superalgebra $\uqlslmn$.

\section{\texorpdfstring{Quantum superalgebra $\uqlslmn$}{Quantum superalgebra Uq(L(slm|n))}} \label{s:4}

Let us start with the necessary information on the affine Lie superalgebra $\hlslmn$. More information can be found in the papers \cite{Leu86, Leu89}.

We denote by $\lslmn$ the loop superalgebra of the Lie superalgebra $\slmn$, by $\tlslmn$ its standard central extension by a one-dimensional center $\bbC c$, and by $\hlslmn$ the Lie superalgebra obtained from $\tlslmn$ by adding a natural derivation $d$. By definition, as a vector space
\begin{equation*}
\hlslmn = \lslmn \oplus \bbC \, c \oplus \bbC \, d, 
\end{equation*}
and we use as the Cartan subsuperalgebra of $\hlslmn$ the subsuperspace\footnote{Here we use the natural embedding of $\slmn$ into $\hlslmn$.} 
\begin{equation*}
\hgothh = \gothh \oplus \bbC \, c \oplus \bbC \, d.
\end{equation*}
We assume that $[c] = [d] = \overline 0$. Further, we define
\begin{equation*}
h_0 =  c - \ \sum_{i = 1}^{\mathclap{M + N - 1}} \ d_i h_i.
\end{equation*}
It is worth to note that
\begin{equation*}
c = \ \sum_{i = 0}^{\mathclap{M + N - 1}}\ d_i h_i,
\end{equation*}
where $d_0 =  1$. It is also convenient to denote
\begin{equation*}
\tgothh = \gothh \oplus \bbC \, c
\end{equation*}
and identify $\gothh$ and $\tgothh$ with the corresponding subspaces of the subsuperalgebra $\hgothh$.

We identify the space $\gothh^*$ with the subspace of $\tgothh^*$ which consists of the elements $\lambda \in \tgothh^*$ satisfying the equation
\begin{equation}
\langle \lambda, \, c \rangle = 0. \label{lambdac}
\end{equation}
Here and henceforth, we mark the elements of this subspace by a tilde. Explicitly, the identification is performed as follows. The element $\widetilde \lambda \in \widetilde \gothh^*$ satisfying equation (\ref{lambdac}) is identified with the element $\lambda \in \gothh^*$ defined by the equations
\begin{equation*}
\langle \lambda, \, h_i \rangle = \langle \widetilde \lambda, \, h_i \rangle, \qquad i = 1, \ldots, M + N - 1.
\end{equation*}
In the opposite direction, given an element $\lambda \in \gothh^*$, identify it with the element 
$\widetilde \lambda \in \widetilde \gothh^*$ determined by the relations
\begin{equation*}
\langle \widetilde \lambda, \, h_0 \rangle = - \ \sum_{i = 1}^{\mathclap{M + N - 1}}  \ \langle \lambda, \, h_i \rangle, \qquad  
\langle \widetilde \lambda, \, h_i \rangle = \langle \lambda, \, h_i \rangle, \quad i = 1, \ldots, M + N - 1.
\end{equation*}
It is clear that the element $\widetilde \lambda$ defined in this way satisfies equation (\ref{lambdac}). 

After all, denote by $\delta$ the element of the space $\hgothh^*$ determined by the equations
\begin{equation*}
\langle \delta, \, h_i \rangle = 0, \quad i = 0, 1, \ldots, M + N - 1, \qquad \langle \delta, \, d \rangle = 1,
\end{equation*}
and define the root $\alpha_0 \in \hgothh^*$ as
\begin{equation*}
\alpha_0 = \delta - \ \sum_{i = 1}^{\mathclap{M + N - 1}} \ \alpha_i = \delta - \alpha_{1, \, M + N}.
\end{equation*}
It is clear that
\begin{equation*}
\delta = \ \sum_{i = 0}^{\mathclap{M + N - 1}} \ \alpha_i.
\end{equation*}
The system of positive roots of the Lie superalgebra $\hlslmn$ is
\begin{multline*}
\widehat \Delta_+ = \{\gamma + n \delta \mid  \gamma \in \Delta_+, \ n \in \bbZ_{\ge 0} \} \\
\cup \{n \delta \mid n \in \bbZ_{>0} \} \cup \{(\delta - \gamma) + n \delta \mid  \gamma \in \Delta_+, \ n \in \bbZ_{\ge 0}\}.
\end{multline*}
The system of negative roots is $\widehat \Delta_- = - \widehat \Delta_+$, and the full root system is
\begin{equation*}
\widehat \Delta = \widehat \Delta_+ \sqcup \widehat \Delta_- 
= \{ \gamma + n \delta \mid \gamma \in \Delta, \ n \in \bbZ \} \cup \{n \delta \mid n \in \bbZ \setminus \{0\} \}.
\end{equation*}
The roots $n \delta$, $n \in \bbZ \setminus \{0\}$ are imaginary, all other roots are real.\footnote{A root $\gamma \in \Delta$ is called imaginary if $k \gamma \in \Delta$ for all $k \in \bbZ \setminus 0$, and real otherwise \cite{Leu86, Leu89}.} We assume that the root $\delta$ is even, then the root $\alpha_0$ is odd. In general we put
\begin{equation*}
[\gamma + n \delta] = [\gamma], \quad \gamma \in \Delta, \ n \in \bbZ, \qquad [n \delta] = \oz, \quad n \in \bbZ \setminus \{0\}.
\end{equation*}

Fix a non-degenerate symmetric bilinear form on $\hgothh$ by the equations
\begin{equation*}
\bil{h_i}{h_j} = a^{}_{i j} \, d^{-1}_j, \qquad 
\bil{h_i}{d} = \delta^{}_{i 0} \, d^{-1}_0, \qquad \bil{d}{d} = 0,
\end{equation*}
where $i, j = 0, 1, \ldots, M + N - 1$. Then, for the corresponding symmetric bilinear form on 
$\hgothh^*$ one has
\begin{equation*}
\bil{\alpha_i}{\alpha_j} = d_i a_{i j}.
\end{equation*}
It follows from this relation that
\begin{equation*}
\bil{\delta}{\gamma} = 0
\end{equation*}
for any $\gamma \in \widehat Q$.

The quantum superalgebra $\uqlslmn$ is a unital associative $\bbC$-superalgebra generated by the elements
\begin{equation*}
q^x, \quad x \in \tgothh, \qquad e_i, \quad f_i, \quad i = 0, 1,\ldots, M + N - 1.
\end{equation*}
The $\bbZ_2$-grading of the superalgebra $\uqlslmn$ is defined on generators as
\begin{equation*}
[q^x] = \overline 0, \quad x \in \tgothh \qquad [e_i] = [f_i] = \left\{ \begin{array}{ll}
\overline 0, & i \ne M, \\[.5em]
\overline 1, & i = 0, M.
\end{array} \right.
\end{equation*}
The abelian group
\begin{equation*}
\widehat Q = \bigoplus_{i = 0}^{M + N - 1} \bbZ \, \alpha_i.
\end{equation*}
is called the root lattice of the Lie superalgebra $\glmn$. Assuming that
\begin{equation*}
q^x \in \uqglmn_0, \qquad e_i \in \uqglmn_{\alpha_i}, \qquad f_i \in \uqglmn_{-\alpha_i},
\end{equation*}
we endow $\uqlslmn$ with a $\widehat Q$-grading. We define the $q$-supercommutator for the quantum superalgebra $\uqlslmn$ in the same way as it was done for the quantum superalgebra $\uqglmn$ in section \ref{s:3}.

The defining relations of the quantum superalgebra $\uqlslmn$ look as \cite{Yam99}
\begin{gather}
q^{\nu c} = 1, \quad \nu \in \bbC, \qquad q^{x_1} q^{x_2} = q^{x_1 + x_2}, \label{djra} \\
q^x e_i \, q^{-x} = q^{\langle \alpha_i, \, x \rangle} e_i, \qquad q^x f_i \, q^{-x} = q^{- \langle \alpha_i, \, x \rangle} f_i, \label{djrb} \\
\ldbr e_i, \, f_j \rdbr = \delta_{i j} \, \frac{q_i^{h_i} - q_i^{- h_i}}{q^{\mathstrut}_i - q_i^{-1}}, \label{djrc} \\
\ldbr e_i, \, e_j \rdbr = 0, \qquad \ldbr f_j, \, f_i \rdbr = 0, \qquad \mbox{ if } \bil {\alpha_i}{\alpha_j} = 0, \label{djrd} \\
\ldbr e_i, \, \ldbr e_i, \, e_{i - 1} \rdbr \rdbr = 0, \qquad \ldbr \ldbr f_{i - 1}, \, f_i \rdbr, \, f_i \rdbr = 0, \qquad \mbox{ if } \bil {\alpha_i}{\alpha_i} \ne 0, \label{djrf1} \\
\ldbr e_i, \, \ldbr e_i, \, e_{i + 1} \rdbr \rdbr = 0, \qquad \ldbr \ldbr f_{i + 1}, \, f_i \rdbr, \, f_i \rdbr = 0, \qquad \mbox{ if } \bil {\alpha_i}{\alpha_i} \ne 0, \label{djrf2} \\
\ldbr \ldbr \ldbr e_{M - 1}, \, e_M \rdbr, \, e_{M + 1} \rdbr, \, e_M \rdbr = 0, \qquad \ldbr f_M, \, \ldbr f_{M + 1}, \, \ldbr f_M, \, f_{M - 1} \rdbr \rdbr \rdbr = 0, \label{djrg} \\*
\ldbr \ldbr \ldbr e_1, \, e_0 \rdbr, \, e_{M + N - 1} \rdbr, \, e_0 \rdbr = 0, \qquad \ldbr f_0, \, \ldbr f_{M + N - 1}, \, \ldbr f_0, \, f_1 \rdbr \rdbr \rdbr = 0. \label{djrh}
\end{gather}
The necessary periodicity with respect to the indices is implied in the relations (\ref{djrf1}) and (\ref{djrf2}). In the case of the quantum superalgebra $\uqlslto$ instead of the relations (\ref{djrg}) and (\ref{djrh}) we have the relations
\begin{gather*}
\ldbr e_0, \, \ldbr e_2, \, \ldbr e_0, \, \ldbr e_2, \, e_1 \rdbr \rdbr \rdbr \rdbr = \ldbr e_2, \, \ldbr e_0, \, \ldbr e_2, \, \ldbr e_0, \, e_1 \rdbr \rdbr \rdbr \rdbr, \\
\ldbr \ldbr \ldbr \ldbr f_2, \, f_1 \rdbr, \, f_0, \rdbr \, f_2 \rdbr, \, f_0 \rdbr = \ldbr \ldbr \ldbr \ldbr f_0, \, f_1 \rdbr, \, f_2, \rdbr \, f_0 \rdbr, \, f_2 \rdbr.
\end{gather*}

From the point of view of quantum integrable systems, it is important that the quantum superalgebra $\uqlslmn$ 
is a Hopf algebra with respect to an appropriately defined co-multiplication $\Delta$, antipode $S$ 
and counit $\varepsilon$. Note that the definition of the quantum superalgebra $\uqlslmn$ is invariant under the replacement of $q$ by $q^{-1}$, in contrast to the definition of the Hopf algebra structure. To eliminate the arising ambiguity, we give the explicit form of the Hopf algebra operations adopted in this paper,
\begin{gather*}
\Delta(q^x) = q^x \otimes q^x, \qquad \Delta(e^{}_i) = e^{}_i \otimes 1 + q_i^{h_i} \otimes e^{}_i, \qquad \Delta(f^{}_i) = f^{}_i \otimes q_i^{- h_i} + 1 \otimes f^{}_i, \\
S(q^x) = q^{- x}, \qquad S(e^{}_i) = - q_i^{- h_i} e^{}_i, \qquad S(f^{}_i) = - f^{}_i \, q_i^{h_i}, \\
\varepsilon(q^h) = 1, \qquad \varepsilon(e^{}_i) = 0, \qquad \varepsilon(f^{}_i) = 0.
\end{gather*}

\section{\texorpdfstring{Cartan--Weyl generators}{Cartan-Weyl generators}} \label{s:5}

Having in mind the $\widehat Q$-grading of the quantum superalgebra $\uqlslmn$, we say that an element $a \in \uqlslmn$ is a root vector corresponding to a root $\gamma \in \widehat \Delta$, if $a \in \uqlslmn_\gamma$. In particular, the generators $e_i$ and $f_i$ are root vectors corresponding to the roots $\alpha_i$ and $- \alpha_i$.

One can construct linearly independent root vectors corresponding to all roots from $\widehat \Delta \subset \widehat Q$, see, for example, the papers \cite{KhoTol93a, KhoTol94, KhoTol94a}. They are called the Cartan--Weyl generators of the quantum superalgebra $\uqlslmn$. If some ordering of roots is chosen, then the appropriately ordered monomials constructed from these vectors form a Poincar\'e--Birkhoff--Witt basis of the quantum superalgebra $\uqlslmn$. To construct the root vectors we follow the procedure, based on a normal ordering of positive roots, proposed by Khoroshkin and Tolstoy \cite{KhoTol93a, KhoTol94, KhoTol94a}.

For the case of a finite dimensional simple Lie superalgebra an order relation $\prec$ is called a normal ordering of positive roots when each nonsimple positive root $\gamma = \alpha + \beta$ is located between $\alpha$ and $\beta$ \cite{KhoTol91}. In our case we assume additionally that all imaginary roots follow each other in an arbitrary order and
\begin{equation}
\alpha + k \delta \prec m \delta \prec (\delta - \beta) + n \delta \label{akd}
\end{equation}
for any $\alpha, \beta \in \Delta_+$, $k, n \in \bbZ_{\ge 0}$ and $m \in \bbZ_{>0}$ \cite{KhoTol93a, KhoTol94, KhoTol94a}.

Assume that some normal ordering of positive roots is chosen.  One says that a pair $(\alpha, \, \beta)$ of positive roots generates a root $\gamma$ if $\gamma = \alpha + \beta$ and $\alpha \prec \beta$. A pair of positive roots $(\alpha, \, \beta)$ generating a root $\gamma$ is said to be minimal if there is no other pair of positive roots $(\alpha', \, \beta')$ generating $\gamma$ such that $\alpha \prec \alpha' \prec \beta' \prec \alpha$.

It is convenient to denote a root vector corresponding to a positive root $\gamma$ by $e_\gamma$, and a root vector corresponding to a negative root $- \gamma$ by $f_\gamma$. Following the papers \cite{KhoTol93a, KhoTol94, KhoTol94a}, we define root vectors by the following inductive procedure. Given a nonsimple positive root $\gamma \in \widehat \Delta_+$, let $(\alpha, \, \beta)$ be a minimal pair of positive roots generating $\gamma$. Then, if the root vectors $e_\alpha$, $e_\beta$ and $f_\alpha$, $f_\beta$ are already constructed, we define the root vectors $e_\gamma$ and $f_\gamma$ as 
\begin{equation*}
e_\gamma \propto \ldbr e_\alpha \, , \, e_\beta \rdbr, \qquad f_\gamma \propto \ldbr f_\beta \, , \, f_\alpha \rdbr.
\end{equation*}
Here $\ldbr \, \cdot \, , \, \cdot \rdbr$ means a natural generalization of the $q$-supercommutator defined in section~\ref{s:3} for the case of the quantum superalgebra $\uqglmn$ to the case of the quantum superalgebra $\uqlslmn$.

We use the normal order of the positive roots of the superalgebra Lie $\hlslmn$ defined as follows, see also \cite{MenTes15}. First we put $\alpha_{i j} + m \delta \prec \alpha_{i j} + n \delta$ if $m < n$, and $(\delta - \alpha_{i j}) + m \delta \prec (\delta - \alpha_{i j}) + n \delta$ if $m > n$. Then $\alpha_{i j} + m \delta \prec \alpha_{k l} + n \delta$ if $i < k$, or if $i = k$ and $j < l$. Further, $(\delta - \alpha_{i j}) + m \delta \prec (\delta - \alpha_{k l}) + n \delta$ if $i > k$, or, if $i = k$ and $j < l$. Finally, we assume that the relation (\ref{akd}) is valid. 

The root vectors are defined inductively. We start with the root vectors corresponding to the roots $\alpha_i$ and $-\alpha_i$, $i = 1, \ldots, M + N - 1$, which we identify with the generators $e_i$ and $f_i$, $i = 1, \ldots, M + N - 1$, of~$\uqlslmn$,
\begin{gather*}
e_{\alpha_i} = e_{\alpha_{i, \, i + 1}} = e_i, \qquad f_{\alpha_i} = f_{\alpha_{i, \, i + 1}} = f_i.
\end{gather*}
The next step is to construct root vectors $e_{\alpha_{i j}}$ and $f_{\alpha_{i j}}$ for all roots $\alpha_{i j} \in \Delta_+$. We assume that
\begin{equation}
e_{\alpha_{i j}} = \ldbr e_{\alpha_{i, \, j - 1}}, \, e_{\alpha_{j - 1, \, j}} \rdbr, \qquad f_{\alpha_{i j}} = \ldbr  f_{\alpha_{j - 1, \, j}}, \, f_{\alpha_{i, \, j - 1}} \rdbr \label{eaijfaij}
\end{equation}
for $j > i + 1$. Further, taking into account that $\alpha_0 = \delta - \alpha_{1, \, M + N}$, we put
\begin{equation}
e_{\delta - \alpha_{1, \, M + N}} = e_0, \qquad f_{\delta - \alpha_{1, \, M + N}} = f_0, \label{edmafdma}
\end{equation}
and define
\begin{equation}
e_{\delta - \alpha_{i, \, M + N}} = \ldbr e_{\alpha_{i - 1, \, i}}, \,  e_{\delta - \alpha_{i - 1, \, M + N}} \rdbr, \qquad f_{\delta - \alpha_{i, \,  M + N}} = \ldbr  f_{\delta - \alpha_{i - 1, \, M + N}}, \, f_{\alpha_{i - 1, \, i}} \rdbr  \label{edmaijfdmaiji}
\end{equation}
for $i > 1$, and
\begin{align}
& e_{\delta - \alpha_{i j}} = \ldbr e_{\alpha_{j, \, j + 1}}, \, e_{\delta - \alpha_{i, \, j + 1}} \rdbr, \qquad f_{\delta - \alpha_{i j}} = \ldbr f_{\delta - \alpha_{i, \, j + 1}}, \, f_{\alpha_{j, \, j + 1}} \rdbr \label{edmaijfdmaijii}
\end{align}
for $j < M + N$. The root vectors corresponding to the roots $\delta$ and $-\delta$ are indexed by the simple roots $\alpha_i$, $i = 1, \ldots, M + N - 1$, and are defined by the relations
\begin{equation}
e'_{\delta; \, \alpha_i} = (-1)^{[\alpha_i]} \ldbr e_{\alpha_i}, \,  e_{\delta - \alpha_i} \rdbr, \qquad f'_{\delta; \, \alpha_i} = (-1)^{[\alpha_i]} \ldbr f_{\delta - \alpha_i}, \, f_{\alpha_i} \rdbr. \label{epdfpd}
\end{equation}
The remaining definitions for the root vectors corresponding to the real roots are
\begin{align}
\intertext{$\boxed{i < M}$}
& e_{\alpha_{i j} + n \delta} = \frac{(-1)^{[\alpha_i]}}{[\bil{\alpha_{i j}}{\alpha_i}]_q} \ldbr e_{\alpha_{i j} + (n - 1)\delta}, \, e'_{\delta; \, \alpha_i} \rdbr,  \label{aijnd}\\
& f_{\alpha_{i j} + n \delta} = \frac{(-1)^{[\alpha_i]}}{[\bil{\alpha_{i j}}{\alpha_i}]_q} \ldbr f'_{\delta; \, \alpha_i} \, f_{\alpha_{i j} + (n - 1)\delta} \rdbr, \\
& e_{(\delta - \alpha_{i j}) + n \delta} = \frac{(-1)^{[\alpha_i]}}{[\bil{\alpha_{i j}}{\alpha_i}]_q} \ldbr e'_{\delta; \, \alpha_i}, \, e_{(\delta - \alpha_{i j}) + (n - 1)\delta} \rdbr, \\
& f_{(\delta - \alpha_{i j}) + n \delta} = \frac{(-1)^{[\alpha_i]}}{[\bil{\alpha_{i j}}{\alpha_i}]_q} \ldbr f_{(\delta - \alpha_{i j}) + (n - 1)\delta}, \, f'_{\delta; \, \alpha_i} \rdbr, \\
\intertext{$\boxed{i \ge M}$}
& e_{\alpha_{i j} + n \delta} = \frac{(-1)^{[\alpha_{i - 1}]}}{[\bil{\alpha_{i j}}{\alpha_{i - 1}}]_q} \ldbr e_{\alpha_{i j} + (n - 1)\delta}, \, e'_{\delta; \, \alpha_{i - 1}} \rdbr, \\
& f_{\alpha_{i j} + n \delta} = \frac{(-1)^{[\alpha_{i - 1}]}}{[\bil{\alpha_{i j}}{\alpha_{i - 1}}]_q} \ldbr f'_{\delta; \, \alpha_{i - 1}}, f_{\alpha_{i j} + (n - 1)\delta} \rdbr, \\
& e_{(\delta - \alpha_{i j}) + n \delta} = \frac{(-1)^{[\alpha_{i -1}]}}{[\bil{\alpha_{i j}}{\alpha_{i - 1}}]_q} \ldbr e'_{\delta; \, \alpha_{i - 1}}, \, e_{(\delta - \alpha_{i j}) + (n - 1)\delta} \rdbr, \\
& f_{(\delta - \alpha_{i j}) + n \delta} = \frac{(-1)^{[\alpha_{i - 1}]}}{[\bil{\alpha_{i j}}{\alpha_{i - 1}}]_q} \ldbr  f_{(\delta - \alpha_{i j}) + (n - 1)\delta}, \, f'_{\delta; \, \alpha_{i - 1}} \rdbr, \label{dmaijnd}
\end{align}
and corresponding to the imaginary roots are
\begin{equation}
e'_{n \delta; \, \alpha_i} = (-1)^{[\alpha_i]} \ldbr e_{\alpha_i + (n - 1)\delta}, \, e_{\delta - \alpha_i} \rdbr, \qquad f'_{n \delta; \, \alpha_i} = (-1)^{[\alpha_i]} \ldbr f_{\delta - \alpha_i}, \, f_{\alpha_i + (n - 1)\delta} \rdbr. \label{epdnfpdn}
\end{equation}
The prime in the notation for the root vectors corresponding to the imaginary roots is explained by the fact that to construct the expression for the universal $R$-matrix one uses another set of root vectors corresponding to these roots. They are introduced by the functional equations
\begin{gather*}
- (q_i^{} - q_i^{-1}) \, e_{\delta; \, \alpha_i}(u) = \log(1 - (q_i^{} - q_i^{-1}) \, e'_{\delta; \, \alpha_i}(u)), \\*
(q_i^{} - q_i^{-1}) \, f_{\delta \, \alpha_i}(u^{-1}) = \log(1 + (q_i^{} - q_i^{-1}) \, f'_{\delta; \, \alpha_i}(u^{-1})),
\end{gather*}
where the generating functions
\begin{align*}
& e'_{\delta; \, \alpha_i}(u) = \sum_{n = 1}^\infty e'_{n \delta; \, \alpha_i} \, u^n, && e_{\delta; \, \alpha_i}(u) 
= \sum_{n = 1}^\infty e_{n \delta; \, \alpha_i} \, u^n, \\
& f'_{\delta; \, \alpha_i}(u^{-1}) = \sum_{n = 1}^\infty f'_{n \delta; \, \alpha_i} \, u^{- n}, && f_{\delta; \, \alpha_i}(u^{-1}) = \sum_{n = 1}^\infty f_{n \delta; \, \alpha_i} \, u^{- n}
\end{align*}
are defined as formal power series in $u$ and $u^{-1}$, respectively.

Let us discuss the difference with the case of the quantum algebra $\uqlslm$, see, for example, the paper \cite{NirRaz19}. First of all note that the expression for the universal $R$-matrix is invariant with respect to the normalization of the Cartan--Weyl generators, see section \ref{s:6}. We use a normalization that is convenient for calculations. The main difference with the case of $\uqlslm$ is related to the presence of isotropic roots. Here we use a hint following from the form of the defining relations for the generators of the second Drindfeld's realization of the quantum superalgebra $\uqlslmn$ \cite{Yam99, Zha14}.

\section{Evaluation modules} \label{s:7}

In applications to the theory of quantum integrable systems, one usually considers families of representations parameterized by a complex parameter called the spectral parameter. We introduce the spectral parameter in the following way. We assume that the quantum superalgebra $\uqlslmn$ is $\bbZ$-graded. It means that $\uqlslmn$ expands into a direct sum
\begin{equation*}
\uqlslmn = \bigoplus_{m \in \bbZ} \uqlslmn_m,
\end{equation*}
where
\begin{equation*}
 \uqlslmn_m \, \uqlslmn_n \subset \uqlslmn_{m + n}.
\end{equation*}
By definition, any element $a \in \uqlslmn$ can be uniquely represented as
\begin{equation*}
a = \sum_{m \in \bbZ} a_m, \qquad a_m \in \uqslmn_m.
\end{equation*}
Here, $a_m = 0$ for all but finitely many values of $m$. Given $\zeta \in \bbC^\times$, we define the grading automorphism $\Gamma_\zeta$ by the equation
\begin{equation*}
\Gamma_\zeta(a) = \sum_{m \in \bbZ} \zeta^m a_m.
\end{equation*}
It is worth noting that
\begin{equation*}
\Gamma_{\zeta_1 \zeta_2} =\Gamma_{\zeta_1} \circ\Gamma_{\zeta_2}
\end{equation*}
for any $\zeta_1, \zeta_2 \in \bbC^\times$. Now, for any representation $\varphi$ of $\uqlslmn$, we define the corresponding family $\varphi_\zeta$ of representations as
\begin{equation*}
\varphi_\zeta = \varphi \circ\Gamma_\zeta.
\end{equation*}

The common way to endow $\uqlslmn$ with a $\bbZ$-grading is to assume that
\begin{equation*}
q^x \in \uqlslmn_0, \qquad e_i \in \uqlslmn_{s_i}, \qquad f_i \in \uqlslmn_{-s_i},
\end{equation*}
where $s_i$ are arbitrary integers. It is clear that for such a $\bbZ$-grading we have
\begin{equation}
\Gamma_\zeta(q^x) = q^x, \qquad\Gamma_\zeta(e_i) = \zeta^{s_i} e_i, \qquad\Gamma_\zeta(f_i) = \zeta^{-s_i} f_i. \label{gzqx}
\end{equation}

To construct representations of $\uqlslmn$, it is common to use the Jimbo's homomorphism $\epsilon$ from the quantum superalgebra $\uqlslmn$ to the quantum superalgebra $\uqglmn$ defined by the equations
\begin{align*}
& \epsilon(q^{\nu h_0}) = q^{- \nu (- d_{M + N} K_{M + N} + d_1 K_1)}, &&  
\epsilon(q^{\nu h_i}) = q^{\nu (K_{i} - d_i d_{i + 1}  K_{i + 1})}, \\
& \epsilon(e_0) = - F_{1,\, M + N} \, q^{d_1 K_1 + d_{M + N} K_{M + N}}, && 
\epsilon(e_i) = E_{i, \,i+1}, \\
& \epsilon(f_0) =  q^{- d_1 K_1 - d_{M + N} K_{M + N}} E_{1, \, M + N}, && 
\epsilon(f_i) = F_{i, \, i + 1},
\end{align*}
where $i$ runs from $1$ to $M + N - 1$. It is clear that if $\pi$ is a representation of the quantum superalgebra $\uqglmn$ then $\pi \circ \epsilon$ is a representation of the quantum superalgebra $\uqlslmn$. We call such representations and the corresponding modules evaluation ones. Starting with the representation $\pi$ the quantum superalgebra $\uqglmn$ defined by equations (\ref{pik}) and (\ref{piepif}), we define a family of evaluation representations
\begin{equation*}
\varphi_\zeta =  \pi \circ \epsilon \circ\Gamma_\zeta
\end{equation*}
of the quantum superalgebra $\uqlslmn$. Using the explicit form of the Jimbo's homomorphism, we obtain
\begin{align}
& \varphi_\zeta(q^{\nu h_0}) =q^{-\nu} \bbE_{1 1} + q^{-\nu} \bbE_{M + N, \, M + N} + \sum_{k = 2}^{M + N - 1} \bbE_{k k}, \label{fzqh0} \\
& \varphi_\zeta(q^{\nu h_i}) = q^\nu \bbE_{i i} +  q^{-\nu d_i d_{i + 1}} \bbE_{i + 1, \, i + 1} + \sum_{\substack{k = 1 \\  k \ne i, \, i + 1}}^{M + N} \bbE_{k k}, \qquad i = 1, \ldots M + N - 1, \label{fzqhi}
\end{align}
and, further,
\begin{align}
& \varphi_\zeta(e_0) = - \zeta^{s_0} q \bbE_{M + N, \, 1}, && \varphi_\zeta(e_i) = \zeta^{s_i} \bbE_{i, \, i + 1}, && i = 1, \ldots, M + N - 1, \label{fzen} \\
& \varphi_\zeta(f_0) = \zeta^{-s_0} q^{-1} \bbE_{1, \, M + N}, && \varphi_\zeta(f_i) = \zeta^{-s_i} \bbE_{i + 1, \, i}, && i = 1, \ldots, M + N - 1. \label{fzfn}
\end{align}
By definition, the representations $\varphi_\zeta$ act on the superspace $\bbC_{M | N}$.

Let $V$ be a $\uqlslmn$-module. The generators $q^x$, $x \in \tgothh$, form an abelian group in $\uqlslmn$. Let a vector $v \in V$ be a common eigenvector for all operators $\varphi(q^x)$, then
\begin{equation*}
q^x v = q^{\langle \mu, \, x \rangle} v
\end{equation*}
for some unique element $\mu \in \tgothh^*$. Using the first relation of (\ref{djra}), we obtain
\begin{equation*}
q^{\nu c} v = q^{\nu \langle \mu, \, c \rangle} v = v
\end{equation*}
for any $\nu \in \bbC$. Therefore, the element $\mu$ satisfies the equation
\begin{equation*}
 \langle \mu, \, c \rangle = 0,
\end{equation*}
and there is a unique element $\lambda \in \gothh^*$ such that $\mu = \widetilde \lambda$. For the definition of $\widetilde \lambda$ see section ~\ref{s:4}. This leads to the following definition. A $\uqlslmn$-module $V$ is said to be a weight module if
\begin{equation*}
V = \bigoplus_{\lambda \in \gothh^*}  V_\lambda,
\end{equation*}
where
\begin{equation*}
V_\lambda = \{v \in V \mid q^x v 
= q^{\langle \widetilde \lambda, \, x \rangle} v \mbox{ for any } x \in \tgothh \}.
\end{equation*}
This means that any vector of $V$ has the form
\begin{equation*}
v = \sum_{\lambda \in \gothh^*} v_\lambda,
\end{equation*}
where $v_\lambda \in V_\lambda$ for any $\lambda \in \gothh^*$, and $v_\lambda = 0$ for all but finitely many of $\lambda$. The space $V_\lambda$ is called the weight space of weight $\lambda$, and a nonzero element of $V_\lambda$ is called a weight vector of weight $\lambda$. One says that $\lambda \in \gothh^*$ is a weight of $V$ if $V_\lambda \ne \{0\}$. Any finite dimensional $\uqlslmn$-module is a weight module.

For any $i = 1, \ldots, M + N - 1$ the weight $\omega_i$ defined by the equation
\begin{equation*}
\langle \omega_i, \, h_j \rangle = \delta_{i j}
\end{equation*}
is called fundamental. One can show that
\begin{equation}
\omega_i = \sum_{k = 1}^{M + N - 1} \alpha_k^{} \, c_{k i} \, d_i, \label{oi}
\end{equation}
where $c_{k i}$ are matrix entries of the matrix $C$ inverse to $B$.

It is not difficult to verify that for the $\uqlslmn$-module defined by equations (\ref{fzqh0})--(\ref{fzfn}) the basis vector $v_i \in \bbC_{M | N}$ is a weight vector of weight
\begin{equation}
\lambda_i = \omega_1 - \sum_{k = 1}^{i - 1} \alpha_k. \label{lkpi}
\end{equation}

\section{\texorpdfstring{Universal $R$-matrix}{Universal R-matrix}} \label{s:6}

Let $\Pi$ be the automorphism of the superalgebra $\uqlslmn \otimes \uqlslmn$ defined by the equation
\begin{equation*}
\Pi (a_1 \otimes a_2) = (-1)^{[a_1] [a_2]} a_2 \otimes a_1.
\end{equation*}
One can show that the mapping 
\begin{equation*}
\Delta' = \Pi \circ \Delta
\end{equation*}
can serve as another comultiplication in $\uqlslmn$ called the opposite comultiplication.

It should be noted that we define the quantum superalgebra $\uqlslmn$ as a $\bbC$-algebra. It can be also defined as a $\bbC[[\hbar]]$-algebra, where $\hbar$ is considered as an indeterminate. In this case there exists an element $\calR \in \uqlslmn^{\otimes 2}$ connecting the comultiplications in the sense that
\begin{equation}
\Delta'(a) = \calR \, \Delta(a) \calR^{-1} \label{dpx}
\end{equation}
for any $a \in \uqlslmn$, and satisfying in $\uqlslmn^{\otimes 3}$ the equations
\begin{equation}
(\Delta \otimes \id)(\calR) = \calR^{(13)} \, \calR^{(23)}, \qquad (\id \otimes \Delta)(\calR) = \calR^{(13)} \, \calR^{(12)}, \label{drrr}
\end{equation}
see \cite{Dri87} for the case of a general quantum algebra, and \cite{GouZhaBra93} for the generalization to the case of quantum superalgebras. For the meaning of the superscripts in the above relations see, for example, appendix A.1 of the paper \cite{NirRaz19}. The element $\calR$ is called the universal $R$-matrix. One can show that it satisfies the universal Yang-Baxter equation
\begin{equation*}
\calR^{(12)} \, \calR^{(13)} \, \calR^{(23)} = \calR^{(23)} \, \calR^{(13)} \, \calR^{(12)}
\end{equation*}
in $\uqlslmn^{\otimes 3}$.

Using a representation $\varphi$ of the quantum superalgebra $\uqlslmn$, one construct the corresponding $R$-operator $R_\varphi(\zeta_1, \zeta_2)$ as
\begin{equation*}
R_\varphi(\zeta_1, \zeta_2) = (\varphi_{\zeta_1} \otimes \varphi_{\zeta_2})(\calR).
\end{equation*}
It satisfies the usual operator Yang--Baxter equation
\begin{equation}
R^{(1 2)}_\varphi(\zeta_1, \zeta_2) R^{(1 3)}_\varphi(\zeta_1, \zeta_3) R^{(2 3)}_\varphi(\zeta_2, \zeta_3) = R^{(2 3)}_\varphi(\zeta_2, \zeta_3) R^{(1 3)}_\varphi(\zeta_1, \zeta_3) R^{(1 2)}_\varphi(\zeta_1, \zeta_2). \label{oybe}
\end{equation}
For the notations used, we again refer to the paper \cite{NirRaz19}.

In our case, the universal $R$-matrix exists only in some restricted sense, see the papers \cite{Tan92, Raz21, Raz21a} for the case of quantum algebras $\uqlslm$, and the discussion below for the case of $\uqlslmn$. Here, an $R$-operator is constructed in the following way. Let $V$ be a finite dimensional $\uqlslmn$-modules, and $\varphi$ be the corresponding representation. Define the $R$-operator $R_\varphi(\zeta_1, \zeta_2)$ as
\begin{equation}
R_\varphi(\zeta_1, \zeta_2) = \rho_\varphi(\zeta_1, \zeta_2) \, (\varphi_{\zeta_1} \otimes \varphi_{\zeta_2})(\calR_{\prec \delta} \, \calR_{\sim \delta} \, \calR_{\succ \delta}) \, K_\varphi. \label{rpipi}
\end{equation}
Here $ \rho_\varphi(\zeta_1, \zeta_2)$ is a scalar normalization factor, $\calR_{\prec \delta}$, $\calR_{\sim \delta}$ and $\calR_{\succ \delta}$ are elements of the superalgebra $\uqlslmn^{\otimes 2}$, while $K_\varphi$ is an element of the algebra $\End(V^{\otimes 2} V)$.

Explicitly, the elements $\calR_{\prec \delta}$ and $\calR_{\succ \delta}$ are the products over the sets of roots $\gamma = \alpha_{i j} + n \delta$ and $\gamma = (\delta - \alpha_{i j}) + n \delta$, respectively, of the $q$-exponentials
\begin{equation*}
\calR_\gamma = \exp_{q_\gamma} \big( - (-1)^{[\gamma]} (q - q^{-1}) a_\gamma^{-1} \, (e_\gamma^{} \otimes f_\gamma^{}) \big),
\end{equation*}
where
\begin{equation*}
q_\gamma = (-1)^{[\gamma]} q^{(\gamma | \gamma)}.
\end{equation*}
It is useful to have in mind that
\begin{equation*}
[\alpha_{i j} + n \delta] = [(\delta - \alpha_{i j}) + n \delta] = [i] + [j], \quad \bil {(\delta - \alpha_{i j}) + n \delta}{(\delta - \alpha_{i j}) + n \delta} = d_i + d_j.
\end{equation*}
The quantities $a_\gamma$ are determined from the equation
\begin{equation*}
\ldbr e_\gamma, \, f_\gamma \rdbr = a_\gamma \frac{q^{h_\gamma} - q^{- h_\gamma}}{q - q^{-1}},
\end{equation*}
where for
$\gamma = \sum_{i = 0}^{M + N - 1} m_i \alpha_i$
we put
\begin{equation*}
h_\gamma = \sum_{i = 0}^{M + N - 1} d_i m_i h_i.
\end{equation*}
In particular, $h_\delta = c$, therefore,
\begin{equation*}
q^{h_\delta} = q^c = 1.
\end{equation*}
The order of the factors in $\calR_{\prec \delta}$ and $\calR_{\succ \delta}$ coincides with the chosen normal order of the roots $\alpha_{i j} + n \delta$ and  $(\delta - \alpha_{i j}) + n \delta$.

The element $\calR_{\sim \delta}$ is defined as\footnote{In fact, in our case $[d_i]_q = d_i$. We write $[d_i]_q$ to present expressions valid for a more general case.}
\begin{equation*}
\calR_{\sim \delta} = \exp \big( - (q - q^{-1}) \sum_{n \in \bbZ_{>0}} \, \sum_{i, j = 1}^{M + N - 1} (-1)^n u_{n i j} [d_i]_q [d_j]_q (e_{n \delta, \, \alpha_i} \otimes f_{n \delta, \, \alpha_j}) \big),
\end{equation*}
where for each $n \in \bbZ_{> 0}$ the quantities $u_{n i j}$ are the matrix entries of the matrix $U_n$ inverse to the matrix $T_n$ with the matrix entries $t_{n i j}$ determined by the equation
\begin{equation}
\ldbr e_{\alpha_i + m \delta}, \, e_{n \delta; \, \alpha_j} \rdbr = [d_j]_q^{-1} \, o_i^n o_j^n t_{n i j}^{} \, e_{\alpha_i + (m + n) \delta}. \label{eaipmd}
\end{equation}
The quantities $o_i$ have the form
\begin{equation*}
o_i = (-1)^{i - 1}, \quad i <  M, \qquad o_i = (-1)^i, \quad i \ge M.
\end{equation*}
In contrast to the case of ordinary quantum algebras, the general form of the quantities $o_i$ for quantum superalgebras is unknown. We found them using matching considerations by selection.

The operator $K_\varphi$ is defined as follow. Let $v, w \in V$ be weight vectors of weights $\lambda, \mu \in \gothh^*$,  respectively. Then we have
\begin{equation}
K_\varphi (v \otimes w) = q^{- \bil {\lambda}{\mu}} v \otimes w.
\label{kvw}
\end{equation}

One can show that it is possible to work with the $R$-operator $R_\varphi(\zeta_1, \zeta_2)$ defined by equation (\ref{rpipi}) as if it were defined by the universal $R$-matrix, satisfying equations (\ref{dpx}) and (\ref{drrr}). In particular, it satisfies the operator Yang--Baxter equation (\ref{oybe}).

\section{\texorpdfstring{Explicit form of $R$-operator}{Explicit form of R-operator}} \label{s:8}

First, construct the expression for the operator $K_\varphi$. Putting $q = 1$ in equations (\ref{ibqa}) and (\ref{ibqb}), we obtain expressions for the matrix entries of the matrix $C = B^{-1}$. Now, using equations (\ref{kvw}), (\ref{lkpi}), and (\ref{oi}), we get
\begin{multline*}
K_\varphi \, v_m \otimes v_n = q^{- \sum_{i, \, j = 1}^{M + N - 1} d_i^{-1} d_j^{-1} \langle \omega_1 - \sum_{k = 1}^{m - 1} \alpha_k, \, h_i \rangle \langle \omega_1 - \sum_{l = 1}^{n - 1} \alpha_l, \, h_j \rangle c_{i j}} \, v_m \otimes v_n \\*
= q^{-(d_1^{-2} c_{1 1} - d_1^{-1} \sum_{k = 1}^{m - 1} \delta_{1 k} - d_1^{-1} \sum_{l = 1}^{n - 1} \delta_{l 1} + \sum_{k = 1}^{m - 1} \sum_{l = 1}^{n - 1} b_{k l})} \, v_m \otimes v_n.
\end{multline*}
One can demonstrate that
\begin{align*}
& K_\varphi \, v_m \otimes v_m =  q^{-(M - N - 1)/(M - N)} v_m \otimes v_m, && m \le M, \\
& K_\varphi \, v_m \otimes v_m =  q^{-(M - N - 1)/(M - N) + 2} \, v_m \otimes v_m, && m > M, \\
& K_\varphi \, v_m \otimes v_n =  q^{-(M - N - 1)/(M - N) + 1} \, v_m \otimes v_n. && m \ne n.
\end{align*}
Using these equations, we obtain the expression
\begin{equation}
K_\varphi = q^{-(M - N - 1)/(M - N)} \Big( \sum_{i = 1}^M \bbE_{i i} \otimes \bbE_{i i} + q^2 \sum_{i = M + 1}^{M + N} \bbE_{i i} \otimes \bbE_{i i} + q \sum_{\substack{i, j, = 1 \\ i \ne j}}^{M + N} \bbE_{i i} \otimes \bbE_{j j} \Big). \label{k}
\end{equation}

Further, it is evident that
\begin{equation*}
\varphi_\zeta(e_{\alpha_{i, \, i + 1}}) = \zeta^{s_{i}} \bbE_{i, \, i + 1}, \qquad \varphi_\zeta(f_{\alpha_{i, \, i + 1}}) = \zeta^{- s_{i}} \bbE_{i + 1, \, i},
\end{equation*}
and, using equations (\ref{eaijfaij}), we obtain
\begin{equation*}
\varphi_\zeta(e_{\alpha_{i j}}) = \zeta^{s_{i j}} \bbE_{i j}, \qquad \varphi_\zeta(f_{\alpha_{i j}}) = \zeta^{-s_{i j}} \bbE_{j i},
\end{equation*}
where
\begin{equation*}
s_{i j} = \sum_{k = i}^{j - 1} s_k.
\end{equation*}
Equations (\ref{edmafdma}) give
\begin{equation*}
\varphi_\zeta(e_{\delta - \alpha_{1, \, M + N}}) = - \zeta^{s - s_{1, \, M + N}} q \, \bbE_{M + N, \, 1}, \qquad 
\varphi_\zeta(f_{\delta - \alpha_{1, \, M + N}}) = \zeta^{- (s - s_{1, \, M + N})} q^{-1} \bbE_{1, \, M + N},
\end{equation*}
and, using (\ref{edmaijfdmaiji}) and (\ref{edmaijfdmaijii}), we come to the equations
\begin{align*}
\intertext{$\boxed{i \le M}$}
& \varphi_\zeta(e_{\delta - \alpha_{i j}}) = - \zeta^{s - s_{i j}} (-1)^{i - 1} q^i \bbE_{j i}, && \varphi_\zeta(f_{\delta - \alpha_{i j}}) = \zeta^{-(s - s_{i j})} (-1)^{-i + 1} q^{-i} \bbE_{i j},\\
\intertext{$\boxed{i > M}$}
& \varphi_\zeta(e_{\delta - \alpha_{i j}}) = - \zeta^{s - s_{i j}} (-1)^i q^{-i + 2 M + 2} \bbE_{j i} && \varphi_\zeta(f_{\delta - \alpha_{i j}}) = \zeta^{-(s - s_{i j})} (-1)^{-i} q^{i - 2 M - 2} \bbE_{i j}.
\end{align*}
It follows from (\ref{epdfpd}) that
\begin{align*}
\intertext{$\boxed{i < M}$}
& \varphi_\zeta(e'_{\delta; \, \alpha_i}) = - \zeta^s (-1)^{i - 1} q^i \big( \bbE_{i i} - q^2 \bbE_{i + 1, \, i + 1} \big), \\
& \varphi_\zeta(f'_{\delta; \, \alpha_i}) = \zeta^{-s} (-1)^{- i + 1} q^{-i} \big( \bbE_{i i} - q^{-2} \bbE_{i + 1, \, i + 1} \big),
\intertext{$\boxed{i = M}$}
& \varphi_\zeta(e'_{\delta; \, \alpha_i}) = - \zeta^s (-1)^M q^M (\bbE_{M M} + \bbE_{M + 1, \, M + 1} \big), \\
& \varphi_\zeta(f'_{\delta; \, \alpha_i}) = \zeta^{-s} (-1)^{-M} q^{-M} (\bbE_{M M} + \bbE_{M + 1, \, M + 1} \big),
\intertext{$\boxed{i > M}$}
& \varphi_\zeta(e'_{\delta; \, \alpha_i}) = - \zeta^s (-1)^i q^{2 M - i + 2} \big( \bbE_{i i} - q^{-2} \bbE_{i + 1, \, i + 1} \big), \\
& \varphi_\zeta(f'_{\delta; \, \alpha_i}) = \zeta^{-s} (-1)^{-i} q^{-2 M + i - 2} \big( \bbE_{i i} - q^2 \bbE_{i + 1, \, i + 1} \big).
\end{align*}
After that all, using (\ref{aijnd})--(\ref{dmaijnd}), we obtain
\begin{align}
\intertext{$\boxed{i < M, \qquad j = i + 1}$}
& \varphi_\zeta(e_{\alpha_{i j} + n \delta}) = \zeta^{s_{i, \, i + 1} + n s} (-1)^{n(i + 1)} q^{n(i + 1)} \bbE_{i, \, i + 1}, \label{eaijnd} \\
& \varphi_\zeta(f_{\alpha_{i j} + n \delta}) = \zeta^{- s_{i, \, i + 1} - n s} (-1)^{- n i} q^{-n (i + 1)} \bbE_{i + 1, \, i}, \\
\intertext{$\boxed{i < M, \qquad j > i + 1}$}
& \varphi_\zeta(e_{\alpha_{i j} + n \delta}) = \zeta^{s_{i j} + n s} (-1)^{n(i + 1)} q^{n i} \bbE_{i j}, \\
& \varphi_\zeta(f_{\alpha_{i j} + n \delta}) = \zeta^{- s_{i j} - n s} (-1)^{- n i} q^{- n i} \bbE_{j i}, \\
\intertext{$\boxed{i \ge M}$}
& \varphi_\zeta(e_{\alpha_{i j} + n \delta}) = \zeta^{s_{i j} + n s} (-1)^{n i} q^{n (2M - i + 1)} \bbE_{i j}, \\
& \varphi_\zeta(f_{\alpha_{i j} + n \delta}) = \zeta^{- s_{i j} - n s} (-1)^{- n(i + 1)} q^{-n (2M - i + 1)} \bbE_{j i}, \label{faijnd}
\end{align}

\begin{align}
\intertext{$\boxed{i < M, \qquad j = i + 1}$}
& \varphi_\zeta(e_{(\delta - \alpha_{i j}) + n \delta}) = - \zeta^{(s - s_{i, \, i + 1}) + n s} (-1)^{(n + 1) i + n + 1} q^{(n + 1) i + n} \bbE_{i + 1, \, i}, \\
& \varphi_\zeta(f_{(\delta - \alpha_{i, \, i + 1}) + n \delta}) = \zeta^{-(s - s_{i, \, i + 1}) - n s} (-1)^{-(n + 1) i + 1} q^{-(n + 1) i - n} \bbE_{i, \, i + 1}, \\
\intertext{$\boxed{i < M, \qquad j > i + 1}$}
& \varphi_\zeta(e_{(\delta - \alpha_{i j}) + n \delta}) = - \zeta^{(s - s_{i j}) + n s} (-1)^{(n + 1)i + n + 1} q^{(n + 1) i} \bbE_{j i}, \\
& \varphi_\zeta(f_{(\delta - \alpha_{i j}) + n \delta}) = \zeta^{-(s - s_{i j}) - n s} (-1)^{-(n + 1) i + 1} q^{-(n + 1) i} \bbE_{i j},
\intertext{$\boxed{i = M}$}
& \varphi_\zeta(e_{(\delta - \alpha_{i j}) + n \delta}) = - \zeta^{(s - s_{i j}) + n s} (-1)^{(n + 1) M + 1} q^{(n + 1) M + n} \bbE_{j i} \\
& \varphi_\zeta(f_{(\delta - \alpha_{i j}) + n \delta}) = \zeta^{-(s - s_{i j}) - n s} (-1)^{-(n + 1)M + n + 1} q^{-(n + 1)M - n} \bbE_{i j},
\intertext{$\boxed{i > M}$}
& \varphi_\zeta(e_{(\delta \alpha_{i j}) + n \delta}) = - \zeta^{(s - s_{i j}) + n s} (-1)^{(n + 1)i} q^{(n + 1)(2 M - i + 1) + 1}\bbE_{j i}, \\
& \varphi_\zeta(f_{(\delta - \alpha_{i j}) + n \delta}) = \zeta^{-(s - s_{i j}) - n s} (-1)^{-(n + 1)i + n} q^{- (n + 1) (2 M - i + 1) - 1} \bbE_{i j},
\end{align}
and (\ref{epdnfpdn}) gives
\begin{align}
\intertext{$\boxed{i < M}$}
& \varphi_\zeta(e'_{n \delta; \, \alpha_i}) = -\zeta^{n s} (-1)^{n i + n} q^{n(i + 1) -1} \big( \bbE_{i i} - q^2 \bbE_{i + 1, \, i + 1} \big), \label{epnd} \\
& \varphi_\zeta(f'_{n \delta; \, \alpha_i}) = \zeta^{- n s} (-1)^{- n i + 1} q^{-n(i + 1) + 1} \big( \bbE_{i i} - q^{-2} \bbE_{i + 1, \, i + 1} \big),
\intertext{$\boxed{i = M}$}
& \varphi_\zeta(e'_{n \delta; \, \alpha_i}) = -\zeta^{n s} (-1)^{n M} q^{n(M + 1) -1} \big( \bbE_{M M} + \bbE_{M + 1, \, M + 1} \big), \\
& \varphi_\zeta(f'_{n \delta; \, \alpha_i}) = \zeta^{- n s} (-1)^{-n M + n + 1} q^{-n(M + 1) + 1} \big( \bbE_{M M} + \bbE_{M + 1, \, M + 1} \big),
\intertext{$\boxed{i > M}$}
& \varphi_\zeta(e'_{n \delta; \, \alpha_i}) = -\zeta^{n s} (-1)^{n i} q^{n(2 M - i + 1) + 1} \big( \bbE_{i i} - q^{-2} \bbE_{i + 1, \, i + 1} \big), \\
& \varphi_\zeta(f'_{n \delta; \, \alpha_i}) = \zeta^{- n s} (-1)^{-n i + n + 1} q^{-n(2 M - i + 1) - 1} \big( \bbE_{i i} - q^2 \bbE_{i + 1, \, i + 1} \big). \label{fpnd}
\end{align}

Now, we find expressions for $\varphi_\zeta(e_{n \delta, \, \alpha_{i j}})$ and $\varphi_\zeta(f_{n \delta, \, \alpha_{i j}})$. Using (\ref{epnd})--(\ref{fpnd}), we obtain for the corresponding generating functions the following expressions
\begin{align*}
\intertext{$\boxed{i < M}$}
& \mathbbm 1 - (q_i^{} - q_i^{-1}) \sum_{n = 1}^\infty \varphi_\zeta(e'_{n \delta, \, \alpha_i}) u^n = \sum_{\substack{k = 1 \\ k \ne i}}^{M + N - 1} \bbE_{k k} \\
& \hspace{9em} {} + \frac{1 + (-1)^i q^{i - 1} \zeta^s u}{1 + (-1)^i q^{i + 1} \zeta^s u} \bbE_{i i} + \frac{1 + (-1)^i q^{i + 3} \zeta^s u}{1 + (-1)^i q^{i + 1} \zeta^s u} \bbE_{i + 1, \, i + 1}, \\
& \mathbbm 1 + (q_i^{} - q_i^{-1}) \sum_{n = 1}^\infty \varphi_\zeta(f'_{n \delta, \, \alpha_i}) u^{-n} = \sum_{\substack{k = 1 \\ k \ne i}}^{M + N - 1} \bbE_{k k} \\
& \hspace{9em} {} + \frac{1 - (-1)^i q^{- i + 1} \zeta^{-s} u^{-1}}{1 - (-1)^i q^{-i - 1} \zeta^{-s} u^{-1}} \bbE_{i i} + \frac{1 - (-1)^i q^{-i - 3} \zeta^{-s} u^{-1}}{1 - (-1)^i q^{- i - 1} \zeta^{-s} u^{-1}} \bbE_{i + 1, \, i + 1},
\intertext{$\boxed{i = M}$}
& \mathbbm 1 - (q_i^{} - q_i^{-1}) \sum_{n = 1}^\infty \varphi_\zeta(e'_{n \delta, \, \alpha_i}) u^n = \sum_{\substack{k = 1 \\ k \ne M}}^{M + N - 1} \bbE_{k k} \\
& \hspace{14em} {} + \frac{1 - (-1)^M q^{M - 1} \zeta^s u}{1 - (-1)^M q^{M + 1} \zeta^s u} (\bbE_{M M} + \bbE_{M + 1, \, M + 1}), \\
& \mathbbm 1 + (q_i^{} - q_i^{-1}) \sum_{n = 1}^\infty \varphi_\zeta(f'_{n \delta, \, \alpha_i}) u^{-n} = \sum_{\substack{k = 1 \\ k \ne M}}^{M + N - 1} \bbE_{k k} \\
& \hspace{14em} {} + \frac{1 + (-1)^M q^{- M + 1} \zeta^{- s} u^{-1}}{1 + (-1)^M q^{- M - 1} \zeta^{- s} u^{-1}} (\bbE_{M M} + \bbE_{M + 1, \, M + 1}),
\intertext{$\boxed{i > M}$}
& \mathbbm 1 - (q_i^{} - q_i^{-1}) \sum_{n = 1}^\infty \varphi_\zeta(e'_{n \delta, \, \alpha_i}) u^n = \sum_{\substack{k = 1 \\ k \ne i}}^{M + N - 1} \bbE_{k k} \\*
& \hspace{7em} {} + \frac{1 - (-1)^i q^{2 M - i + 3} \zeta^s u}{1 - (-1)^i q^{2 M - i + 1} \zeta^s u} \bbE_{i i} + \frac{1 - (-1)^i q^{2 M - i - 1} \zeta^s u}{1 - (-1)^i q^{2 M - i + 1} \zeta^s u} \bbE_{i + 1, \, i + 1}, \\
& \mathbbm 1 + (q_i^{} - q_i^{-1}) \sum_{n = 1}^\infty \varphi_\zeta(f'_{n \delta, \, \alpha_i}) u^{-n} = \sum_{\substack{k = 1 \\ k \ne i}}^{M + N - 1} \bbE_{k k} \\
& \hspace{7em} {} + \frac{1 + (-1)^i q^{-2 M + i - 3} \zeta^{- s} u^{-1}}{1 + (-1)^i q^{-2 M + i - 1} \zeta^s u^{-1}} \bbE_{i i} + \frac{1 + (-1)^i q^{- 2 M + i + 1} \zeta^s u^{-1}}{1 + (-1)^i q^{- 2 M + i - 1} \zeta^{- s} u^{-1}} \bbE_{i + 1, \, i + 1},
\end{align*}
where
\begin{equation*}
\mathbbm 1 = \sum_{i, j = 1}^{M + N} \bbE_{i i} \otimes \bbE_{j j}
\end{equation*}
is the identity element of the algebra $\End(\bbC_{M | N})$, and come to the equations
\begin{align*}
\intertext{$\boxed{i < M}$}
& \varphi_\zeta(e_{n \delta; \, \alpha_i}) = - \zeta^{n s} (-1)^{n i + n} q^{n i} \frac{[n]_q}{n} \big( \bbE_{i i} - q^{2 n} \bbE_{i + 1, \, i + 1} \big), \\
& \varphi_\zeta(f_{n \delta; \, \alpha_i}) = \zeta^{- n s} (-1)^{- n i + 1} q^{-n i}  \frac{[n]_q}{n} \big( \bbE_{i i} - q^{-2 n} \bbE_{i + 1, \, i + 1} \big),
\intertext{$\boxed{i = M}$}
& \varphi_\zeta(e_{n \delta; \, \alpha_i}) = - \zeta^{n s} (-1)^{n M} q^{n M}  \frac{[n]_q}{n} \big(\bbE_{M M} + \bbE_{M + 1, \, M + 1} \big), \\
& \varphi_\zeta(f_{n \delta; \, \alpha_i}) = \zeta^{- n s} (-1)^{-n M + n + 1} q^{-n M}  \frac{[n]_q}{n} \big(\bbE_{M M} + \bbE_{M + 1, \, M + 1} \big),
\intertext{$\boxed{i > M}$}
& \varphi_\zeta(e_{n \delta; \, \alpha_i}) = - \zeta^{n s} (-1)^{n i} q^{n(2 M - i + 2)}  \frac{[n]_q}{n} \big( \bbE_{i i} - q^{-2 n} \bbE_{i + 1, \, i + 1} \big), \\
& \varphi_\zeta(f_{n \delta; \, \alpha_i}) = \zeta^{- n s} (-1)^{-n i + n + 1} q^{-n(2 M - i + 2)}  \frac{[n]_q}{n} \big( \bbE_{i i} - q^{2 n} \bbE_{i + 1, \, i + 1} \big)
\end{align*}

Applying to equation (\ref{eaipmd}) the representation $\varphi_\zeta$, we find
\begin{equation*}
t_{n i j} = o_i^n o_j^n \frac{[n b_{i j}]_q}{n} = o_i^n o_j^n \frac{[n]_q}{n} [b_{i j}]_{q^n}.
\end{equation*}
It is clear that to find the inverse of $T_n$ it is sufficient to find the inverse of the matrix
\begin{equation}
B_q = ([b_{ij}]_q)_{i, j = 1}^{M + N - 1}. \label{bq}
\end{equation}
The explicit form of the matrix entries of $B_q^{-1}$ is obtained in the appendix \ref{a:2}.

Now, we get
\begin{align*}
& (\varphi_{\zeta_1} \otimes \varphi_{\zeta_2}) \big( - (q - q^{-1}) \sum_{n = 1}^\infty \sum_{i, \, j = 1}^{M + N - 1} U_{n i j} \, e_{n \delta, \, \alpha_i} \otimes f_{n \delta, \, \alpha_j} \big) \\*
& \hspace{4em} {} = - \sum_{n = 1}^\infty \frac{q^{n (M - N - 1)} - q^{- n (M - N - 1)}}{[M - N]_{q^n}} \frac{\zeta^{n s}_{1 2}}{n} (\mathbbm 1 \otimes \mathbbm 1) \\
& \hspace{5em} {} - \sum_{n = 1}^\infty (q^{- 2 n} - 1) \frac{\zeta^{n s}_{1 2}}{n} \sum_{\substack{i, \, j = 1 \\ i < j}}^{M + N} \bbE_{i i} \otimes \bbE_{j j} - \sum_{n = 1}^\infty (1 - q^{2 n}) \frac{\zeta^{n s}_{1 2}}{n} \sum_{\substack{i, \, j = 1 \\ i > j}}^{M + N} \bbE_{i i} \otimes \bbE_{j j} \\
& \hspace{18em} {} - \sum_{n = 1}^\infty (q^{- 2 n} - q^{2 n}) \frac{\zeta^{n s}_{1 2}}{n} \sum_{i = M + 1}^{M + N} \bbE_{i i} \otimes \bbE_{i i}.
\end{align*}
Introducing the transcendental function
\begin{equation*}
F_m(\zeta) = \sum_{n = 1}^\infty \frac{1}{[m]_{q^n}} \frac{\zeta^n}{n},
\end{equation*}
and performing the summations over $n$, we obtain
\begin{align*}
& (\varphi_{\zeta_1} \otimes \varphi_{\zeta_2}) \big( - (q - q^{-1}) \sum_{n = 1}^\infty \sum_{i, \, j = 1}^{M + N - 1} (u_n)_{i j} \, e_{n \delta, \, \alpha_i} \otimes f_{n \delta, \, \alpha_j} \big) \\*
& \hspace{4em} {} = - (F_{M - N}(q^{M - N - 1} \zeta^s_{1 2}) - F_{M - N}(q^{-(M - N - 1)} \zeta^s_{1 2})) \sum_{i, \, j = 1}^{M - N} \bbE_{i i} \otimes \bbE_{j j} \\*
& \hspace{5em} {} + \log \frac{1 - q^{- 2} \zeta^s_{1 2}}{1 - \zeta^s_{1 2}} \sum_{\substack{i, \, j = 1 \\ i < j}}^{M + N} \bbE_{i i} \otimes \bbE_{j j} + \log \frac{1 - \zeta^s_{1 2}}{1 - q^2 \zeta^s_{1 2}} \sum_{\substack{i, \, j = 1 \\ i > j}}^{M + N} \bbE_{i i} \otimes \bbE_{j j} \\
& \hspace{20em} + \log \frac{1 -q^{- 2} \zeta^s_{1 2}}{1 - q^2 \zeta^s_{1 2}} \sum_{i = M + 1}^{M + N} \bbE_{i i} \otimes \bbE_{i i} . 
\end{align*}
After all, we see that
\begin{multline}
(\varphi_{\zeta_1} \otimes \varphi_{\zeta_2}) (R_{\sim \delta}) = \rme^{- F_{M - N}(q^{M - N - 1} \zeta^s_{1 2}) + F_{M - N}(q^{-(M - N - 1)} \zeta^s_{1 2})} \\*
\times \bigg[ \sum_{i = 1}^M \bbE_{i i} \otimes \bbE_{i i} + \frac{1 -q^{- 2} \zeta^s_{1 2}}{1 - q^2 \zeta^s_{1 2}} \sum_{i = M + 1}^{M + N} \bbE_{i i} \otimes \bbE_{i i} \\* + \frac{1 - q^{- 2} \zeta^s_{1 2}}{1 - \zeta^s_{1 2}} \sum_{\substack{i, \, j = 1 \\ i < j}}^{M + N} \bbE_{i i} \otimes \bbE_{j j} + \frac{1 - \zeta^s_{1 2}}{1 - q^2 \zeta^s_{1 2}} \sum_{\substack{i, \, j = 1 \\ i > j}}^{M + N} \bbE_{i i} \otimes \bbE_{j j} \bigg]. \label{rsd}
\end{multline}

Proceed to the factors $(\varphi_{\zeta_1} \otimes \varphi_{\zeta_2}) (\calR_{\prec \delta})$ and $(\varphi_{\zeta_1} \otimes \varphi_{\zeta_2}) (\calR_{\succ \delta})$. First of all we determine that
\begin{equation*}
(\varphi_{\zeta_1} \otimes \varphi_{\zeta_2})(\ldbr e_{\alpha_{i j}  + n \delta}, \, f_{\alpha_{i j} + n \delta} \rdbr) = (-1)^n d_i \frac{(\varphi_{\zeta_1} \otimes \varphi_{\zeta_2})(q^{\sum_{k = i}^{j - 1} d_k h_k} - q^{- \sum_{k = i}^{j - 1} d_k h_k})}{q - q^{-1}}.
\end{equation*}
It follows that
\begin{equation*}
a_{\alpha_{i j} + n \delta} = (-1)^n d_i.
\end{equation*}
Using equations (\ref{eaijnd}) and (\ref{faijnd}), we find that
\begin{equation*}
(\varphi_{\zeta_1} \otimes \varphi_{\zeta_2}) (\calR_{\alpha_{i j} + n \delta}) = \exp_{(-1)^{[i] + [j]} q_i q_j} \big( -\zeta_{1 2}^{s_{i j} + n s} (q - q^{-1}) (-1)^{[j]} (\bbE_{i j} \otimes \bbE_{j i}) \big).
\end{equation*}
Since
\begin{equation*}
(\bbE_{i j} \otimes \bbE_{j i})^k = 0
\end{equation*}
for all $1 \le i < j \le M + N$ and $k > 1$, we obtain
\begin{equation*}
(\varphi_{\zeta_1} \otimes \varphi_{\zeta_2}) (\calR_{\alpha_{i j} + n \delta}) = \mathbbm 1 \otimes \mathbbm 1 - \zeta_{1 2}^{s_{i j} + n s} (q - q^{-1}) (-1)^{[j]} (\bbE_{i j} \otimes \bbE_{j i}).
\end{equation*}
Taking into account that
\begin{equation*}
(\bbE_{i j} \otimes \bbE_{j i})(\bbE_{k m} \otimes \bbE_{m k}) = 0
\end{equation*}
for all $1 \le i < j \le M + N$ and $1 \le k < m \le M + N$, we see that the factors  $(\varphi_{\zeta_1} \otimes \varphi_{\zeta_2}) (\calR_{\alpha_{i j} + n \delta})$ of the operator $(\varphi_{\zeta_1} \otimes \varphi_{\zeta_2}) (\calR_{\prec \delta})$ can be taken in an arbitrary order and obtain the expression
\begin{equation}
(\varphi_{\zeta_1} \otimes \varphi_{\zeta_2}) (R_{\prec \delta}) = \mathbbm 1 \otimes \mathbbm 1 - \frac{q - q^{-1}}{1 - \zeta^s} \sum_{\substack{i, \, j = 1 \\ i < j}}^{M + N} (-1)^{[j]} \, \zeta_{1 2}^{s_{i j}} \, \bbE_{i j} \otimes \bbE_{j i}. \label{rld}
\end{equation}
In a similar way we come to the equation
\begin{equation}
(\varphi_{\zeta_1} \otimes \varphi_{\zeta_2}) (R_{\succ \delta}) = \mathbbm 1 \otimes \mathbbm 1 - \frac{q - q^{-1}}{1 - \zeta^s} \sum_{\substack{i, \, j = 1 \\ i > j}}^{M + N} (-1)^{[j]} \, \zeta_{1 2}^{s - s_{j i}} \, \bbE_{i j} \otimes \bbE_{j i}. \label{rgd}
\end{equation}

Finally, using equations (\ref{rld}), (\ref{rsd}), (\ref{rgd}) and (\ref{k}) we obtain the following expression for the $R$-operator
\begin{multline*}
R(\zeta_1 | \zeta_2) = \sum_{i = 1}^M \bbE_{i i} \otimes \bbE_{i i} + \frac{q^2 (1 - q^{-2} \zeta_{1 2}^s)}{1 - q^2 \zeta_{1 2}^s} \sum_{i = M + 1}^{M + N} \bbE_{i i} \otimes \bbE_{i i} \\*
+ \frac{q (1 - \zeta_{1 2}^s)}{1 - q^2 \zeta_{1 2}^s} \sum_{\substack{i, j = 1 \\ i \ne j}}^{M + N} \bbE_{i i} \otimes \bbE_{j j} + \frac{1 - q^2}{1 - q^2 \zeta_{1 2}^s} \sum_{\substack{i, j = 1 \\ i < j}}^{M + N} (-1)^{[j]} \, \zeta_{1 2}^{s_{i j}} \, \bbE_{i j} \otimes \bbE_{j i}  \\*
+ \frac{1 - q^2}{1 - q^2 \zeta_{1 2}^s} \sum_{\substack{i, j = 1 \\ i > j}}^{M + N} (-1)^{[j]} \, \zeta_{1 2}^{s - s_{j i}} \, \bbE_{i j} \otimes \bbE_{j i},
\end{multline*}
where we assumed that the normalization factor has the form
\begin{equation}
\rho(\zeta_1 | \zeta_2) =  q^{-(M - N - 1)/(M - N)} \rme^{- F_{M - N}(q^{M - N - 1} \zeta^s_{1 2}) + F_{M - N}(q^{-(M - N - 1)} \zeta^s_{1 2})}.
\end{equation}
One can see that up to normalization and arbitrariness of the grading, we obtain the Perk--Schultz $R$-operator \cite{PerSch81}.

\section{Conclusion}

In this paper, we have studied a generalization of the Khoroshkin--Tolstoy approach to the construction of a universal $R$-matrix, which is necessary in the case of quantum superalgebras. In fact, our consideration can be seen as a refinement of the general procedure outlined in the papers \cite{KhoTol93a, KhoTol94, KhoTol94a}. Using the obtained results, we have constructed an explicit expression for the $R$-operator for the case of a vector evaluation representation of the quantum superalgebra $\uqlslmn$. In the future, we plan to construct expressions for other integrability objects and obtain for them the corresponding functional relations.

\vspace{1em}

\subsection*{Acknowledgments}

This work was supported in part by the RFBR grant \#~20-51-12005. The author is grateful to H.~ Boos, F.~ G\"ohmann, A.~ Kl\"umper, and Kh.~S.~Nirov in collaboration with whom some important results, used in this paper, were previously obtained, for useful discussions.

\subsection*{Conflict of interest}

The author declare that he has no conflicts of interest.

\appendix

\section{Superalgebras and Lie superalgebras} \label{a:1}

It is common to call a $\bbZ_2$-graded vector space
\begin{equation*}
V = V_\oz \oplus V_\oo
\end{equation*}
a superspace.  The elements of the subspaces $V_\oz$ and $V_{\overline 1}$ are said to be even and odd, respectively. An element belonging to $V_\oz$ or $V_\oo$ is said to be homogeneous. The parity of a homogeneous element $v$, denoted as $[v]$, is $\overline 0$ or $\overline 1$ according to whether it is in $V_\oz$ or $V_\oo$.\footnote{Throughout the paper we use the following convenient convention. If $v$ is an element of a superspace and $[v]$ appears in some formula or expression, then $v$ is assumed to be homogeneous.} For a finite dimensional superspace $V$, the pair $(\dim V_\oz, \, \dim V_\oo)$ is called the dimension of $V$. The tensor product $V \otimes W$ of two superspaces $V$ and $W$ is considered as a superspace with
\begin{equation*}
(V \otimes W)_\alpha = \bigoplus_{\beta + \gamma = \alpha} V_\beta \otimes W_\gamma.
\end{equation*}

If an algebra $A$ considered as a vector space is a superspace, it is called a superalgebra if
\begin{equation*}
A_\alpha A_\beta \subset A_{\alpha + \beta}, \qquad \alpha, \beta \in \bbZ_2.
\end{equation*}
It follows that for any two homogeneous elements $a_1$ and $a_2$ the product $a_1 a_2$ is homogeneous and
\begin{equation*}
[a_1 a_2] = [a_1] + [a_2].
\end{equation*}
A superalgebra with an associative multiplication is called an associative superalgebra. The tensor product $A \otimes B$ of two associative superalgebras $A$ and $B$ is considered as an associative superalgebra with the multiplication defined by the equation
\begin{equation*}
(a_1 \otimes b_1)(a_2 \otimes b_2) = (-1)^{[b_1][a_2]} (a_1 a_2) \otimes (b_1 b_2).
\end{equation*}
The multiplication in the tensor product of any number of associative superalgebras is defined recursively.

Let $V$ be a superspace. The algebra $\End(V))$ becomes an associative superalgebra if one assumes that $a \in (\End(V))_\oz$ when $a V_\alpha \subset V_\alpha$, and $a \in (\End(V)_\oo$ when $a V_\alpha \subset V_{\alpha + \oo}$ for $\alpha \in \bbZ_2$.

A Lie superalgebra $\gothg$ is a superalgebra with a Lie bracket $[\, \cdot \, , \, \cdot \,]$ satisfying the conditions
\begin{gather*}
[x, \, y] = (-1)^{[x] [y]} [y, \, x], \\
(-1)^{[x] [z]}[x, \, [y, \, z]] + (-1)^{[y] [x]}[y, \, [z, \, x]] + (-1)^{[z] [y]}[z, \, [x, \, y]] = 0.
\end{gather*}
Here and throughout the article, it is assumed that
\begin{equation*}
(-1)^\oz =  1, \qquad (-1)^\oo =  - 1.
\end{equation*}
It is clear that
\begin{equation*}
[\gothg_\oz, \, \gothg_\oz] \subset \gothg_\oz, \qquad [\gothg_\oz, \, \gothg_\oo] \subset \gothg_\oo.
\end{equation*}
It means that $\gothg_\oz$ is a Lie algebra, and $\gothg_\oo$ is a $\gothg_\oz$-module.

Given an associative superalgebra $A$, one defines the supercommutator on homogeneous elements of $A$ by
\begin{equation*}
[a_1, \, a_2] =  a_1 a_2 - (-1)^{[a_1] [a_2]} a_2 a_1
\end{equation*}
and then extends the definition by linearity to all elements of $A$. With respect to the supercommutator, the algebra $A$ is a Lie superalgebra. In this way, for any superspace $V$, starting with the superalgebra $\End(V)$ we obtain a Lie superalgebra called the general linear superalgebra and denoted by $\mathfrak{gl}(V)$. 

\section{\texorpdfstring{Inverse of $B_q$}{Inverse of Bq}} \label{a:2}

In this appendix we find an explicit expression for the matrix entries of the matrix $C_q = B_q^{-1}$, where the matrix $B_q$ is defined by equation (\ref{bq}). The matrix $B_q$ is tridiagonal and to find its inverse we use the results of the paper \cite{Usm94}. Let $M = (m_{i j})_{i, j = 1}^L$ be an arbitrary tridiagonal matrix of size $L$. Introduce the notation  $\alpha_i = m_{i, \, i - 1}$, $i = 2, \ldots, L$, $\beta_i = m_{i i}$, $i = 1, \ldots, L$, and $\gamma_i = m_{i, \, i + 1}$, $i = 1, \ldots, L - 1$. Denote also the principal minors of $M$ as
\begin{equation*}
\theta_k = \det (m_{i j})_{i, j = 1}^k, \qquad \varphi_k = \det (m_{i j})_{i, j = k}^L.
\end{equation*}
Assuming that
\begin{equation*}
\theta_{-1} = 0, \qquad \theta_0 = 1, \qquad \varphi_{L + 1} = 1, \qquad \varphi_{L + 2} = 0,
\end{equation*}
one can prove the validity of the recurrence relations 
\begin{align}
& \theta_i = \beta_i \theta_{i - 1} - \alpha_i \gamma_{i - 1} \theta_{i - 2}, \label{thetai} \\
& \varphi_i = \beta_i \varphi_{i + 1} - \gamma_i \alpha_{i + 1} \varphi_{i + 2}, \label{phii} 
\end{align}
and demonstrate that
\begin{equation}
m^{-1}_{i j} = \left\{ \begin{array}{ll}
(-1)^{i + j} \gamma_i \gamma_{i + 1} \ldots \gamma_{j - 1} \theta_{i - 1} \varphi_{j + 1} / \theta_N, & i < j, \\[.5em]
\theta_{i - 1} \varphi_{i + 1} / \theta_N, & i = j, \\[.5em]
(-1)^{i + j} \alpha_{j + 1} \alpha_{j + 2} \ldots \alpha_i \theta_{j - 1} \varphi_{i + 1} / \theta_N, & i > j.
\end{array} \right. \label{miij}
\end{equation}

The matrix $B_q$ is symmetric, therefore it is enough to find the matrix elements $B_q^{-1}$ only for $i \le j$. We have
\begin{equation*}
\alpha_i = \left \{ \begin{array}{rl}
-1, & i \le M, \\[.5em]
1, & i > M
\end{array} \right. , \qquad
\beta_i = \left \{ \begin{array}{rl}
[2]_q, & i < M, \\[.5em]
0, \, & i = M, \\[.5em]
-[2]_q, & i > M
\end{array} \right., \qquad
\gamma_i = \left \{ \begin{array}{rl}
-1, & i < M, \\[.5em]
1, & i \ge M
\end{array} \right. .
\end{equation*}
It is evident that
\begin{equation*}
\prod_{k = i}^{j - 1} \gamma_k = \left\{ \begin{array}{ll}
(-1)^{j - i}, \quad & j \le M, \\[.5em]
(-1)^{M - i + 1}, \quad & i \le M, \quad j > M, \\[.5em]
1, \quad & i > M, \quad j > M. 
\end{array} \right.
\end{equation*}
Using the recurrence relations (\ref{thetai}) and (\ref{phii}), we find
\begin{align*}
& \theta_i = \left \{ \begin{array}{ll}
[i + 1]_q, & i < M, \\[.5em]
(-1)^{M + i + 1}[2 M - i - 1]_q, & i \ge M,
\end{array} \right. \\
& \varphi_i = \left \{ \begin{array}{ll}
(-1)^N [M - N - i + 1]_q, & i \le M, \\[.5em]
(-1)^{M + N - i} [M + N - i + 1]_q, & i > M.
\end{array} \right.
\end{align*}
It follows from (\ref{miij}) that
\begin{equation}
c_{q i j} = \left \{
\begin{array}{ll}
\hspace{.9em} [i]_q [M - N - j]_q / [M - N]_q, & j < M, \\[.5em]
-[i]_q [N]_q / [M - N]_q, & j = M, \\[.5em]
-[i]_q [M + N - j]_q / [M - N]_q, & i < M, \quad j > M, \\[.5em]
- [M]_q [M + N - j]_q / [M - N]_q, & i = M, \quad j > M, \\[.5em]
- [2 M - i]_q [M + N - j]_q / [M - N]_q & i > M
\end{array}
\right. \label{ibqa}
\end{equation}
for $i \le j$ and
\begin{equation}
c_{q i j} = c_{q j i} \label{ibqb}
\end{equation}
for $i > j$.

The corresponding expression for the matrix entries of the matrix $C = B^{-1}$ can be obtained by replacing all $q$-numbers by usual numbers.

\providecommand{\href}[2]{#2}


\begin{thebibliography}{10}

\bibitem{BooGoeKluNirRaz14a}

H.~Boos, F.~G{\"o}hmann, A.~Kl\"umper, Kh.~S. Nirov, and A.~V. Razumov,
  \emph{Universal ${R}$-matrix and functional relations},
  \href{http://dx.doi.org/10.1142/S0129055X14300052}{Rev. Math. Phys.}
  \textbf{26} (2014), 1430005 (66pp),
  \href{http://arxiv.org/abs/1205.1631}{{\tt arXiv:1205.1631 [math-ph]}}.

\bibitem{Raz21}

A.~V. Razumov, \emph{$\ell$-weights and factorization of transfer operators},
  \href{http://dx.doi.org/10.1134/S0040577921080092}{Theor. Math. Phys.}
  \textbf{208} (2021), 1116--1143, \href{http://arxiv.org/abs/2103.16200}{{\tt
  arXiv:2103.16200 [math-ph]}}.

\bibitem{Raz21a}

A.~V. Razumov, \emph{Quantum groups and functional relations for arbitrary
  rank}, \href{http://dx.doi.org/10.1016/j.nuclphysb.2021.115517}{Nucl. Phys.
  B} \textbf{971} (2021), 115517 (51pp.),
  \href{http://arxiv.org/abs/2104.12603}{{\tt arXiv:2104.12603 [math-ph]}}.

\bibitem{BazLukZam96}

V.~V. Bazhanov, S.~L. Lukyanov, and A.~B. Zamolodchikov, \emph{Integrable
  structure of conformal field theory, quantum {K}d{V} theory and thermodynamic
  {B}ethe ansatz}, \href{http://dx.doi.org/10.1007/BF02101898}{Commun. Math.
  Phys.} \textbf{177} (1996), 381--398,
  \href{http://arxiv.org/abs/hep-th/9412229}{{\tt arXiv:hep-th/9412229}}.

\bibitem{BazLukZam97}

V.~V. Bazhanov, S.~L. Lukyanov, and A.~B. Zamolodchikov, \emph{Integrable
  structure of conformal field theory {II}. {Q}-operator and {DDV} equation},
  \href{http://dx.doi.org/10.1007/s002200050240}{Commun. Math. Phys.}
  \textbf{190} (1997), 247--278,
  \href{http://arxiv.org/abs/hep-th/9604044}{{\tt arXiv:hep-th/9604044}}.

\bibitem{BazLukZam99}

V.~V. Bazhanov, S.~L. Lukyanov, and A.~B. Zamolodchikov, \emph{Integrable
  structure of conformal field theory {III}. {T}he {Y}ang--{B}axter relation},
  \href{http://dx.doi.org/10.1007/s002200050531}{Commun. Math. Phys.}
  \textbf{200} (1999), 297--324,
  \href{http://arxiv.org/abs/hep-th/9805008}{{\tt arXiv:hep-th/9805008}}.

\bibitem{Dri85}

V.~G. Drinfeld, \emph{{H}opf algebras and the quantum {Y}ang-–{B}axter
  equation (in {R}ussian)}, Dokl. Akad. Nauk SSSR \textbf{283} (1985),
  1060--1064.

\bibitem{Jim85}

M.~Jimbo, \emph{A $q$-difference analogue of {$\mathrm U(\mathfrak g)$} and the
  {Y}ang-{B}axter equation}, \href{http://dx.doi.org/10.1007/BF00704588}{Lett.
  Math. Phys.} \textbf{10} (1985), 63--69.

\bibitem{Dri87}

V.~G. Drinfeld, \emph{Quantum groups}, Proceedings of the International
  Congress of Mathematicians, Berkeley, 1986 (A.~E. Gleason, ed.), vol.~1,
  American Mathematical Society, Providence, 1987, pp.~798--820.

\bibitem{Ros89}

M.~Rosso, \emph{An analogue of {P.B.W.} theorem and the universal {$R$}-matrix
  for {$U_h sl(N+1)$}}, \href{http://dx.doi.org/10.1007/BF01219200}{Commun.
  Math. Phys.} \textbf{124} (1989), 307--318.

\bibitem{KirRes90}

A.~N. Kirillov and N.~Reshetikhin, \emph{$q$-{W}eyl group and a multiplicative
  formula for universal {$R$}-matrices},
  \href{http://dx.doi.org/10.1007/BF02097710}{Commun. Math. Phys.} \textbf{134}
  (1990), 421--431.

\bibitem{LevSoi91}

S.~Z. Levendorskii and Ya.~S. Soibelman, \emph{Quantum {W}eyl group and
  multiplicative formula for the ${R}$-matrix of a simple {L}ie algebra},
  \href{http://dx.doi.org/10.1007/BF01079599}{Funct. Anal. Appl.} \textbf{25}
  (1991), 143--145.

\bibitem{LevSoiStu93}

S.~Levendorskii, Ya. Soibelman, and V.~Stukopin, \emph{The quantum {W}eyl group
  and the universal quantum {$R$}-matrix for affine {L}ie algebra
  {$A_1^{(1)}$}}, \href{http://dx.doi.org/10.1007/BF00777372}{Lett. Math.
  Phys.} \textbf{27} (1993), 253--264.

\bibitem{Dam98}

I.~Damiani, \emph{La {$R$}-matrice pour les alg{\`e}bres quantiques de type
  affine non tordu},
  \href{http://dx.doi.org/10.1016/S0012-9593(98)80104-3}{Ann. Sci. {\'E}cole
  Norm. Sup.} \textbf{31} (1998), 493--523.

\bibitem{Dam00}

I.~Damiani, \emph{The {$R$}-matrix for (twisted) affine quantum algebras},
  Representations and Quantizations (Shanghai, 1998), China Higher Education
  Press, Beijing, 2000, pp.~89--144, \href{http://arxiv.org/abs/1111.4085}{{\tt
  arXiv:1111.4085 [math.QA]}}.

\bibitem{TolKho92}

V.~N. Tolstoy and S.~M. Khoroshkin, \emph{The universal {$R$}-matrix for
  quantum untwisted affine {L}ie algebras},
  \href{http://dx.doi.org/10.1007/BF01077085}{Funct. Anal. Appl.} \textbf{26}
  (1992), 69--71.

\bibitem{Lus93}

G.~Lusztig, \emph{Introduction to quantum groups}, Birkh\"auser, Boston, 1993.

\bibitem{KhoTol93}

S.~M. Khoroshkin and V.~N. Tolstoy, \emph{On {D}rinfeld's realization of
  quantum affine algebras},
  \href{http://dx.doi.org/10.1016/0393-0440(93)90070-U}{J. Geom. Phys.}
  \textbf{11} (1993), 445--452.

\bibitem{Yam94}

H.~Yamane, \emph{Quantized enveloping algebras associated with simple {L}ie
  superalgebras and their universal ${R}$-matrices},
  \href{http://dx.doi.org/10.2977/prims/1195166275}{Publ. RIMS. Kyoto Univ.}
  \textbf{30} (1994), 15--87.

\bibitem{Yam99}

H.~Yamane, \emph{On defining relations of affine {L}ie superalgebras and affine
  quantized universal enveloping algebras},
  \href{http://dx.doi.org/10.2977/prims/1195143607}{Publ. RIMS. Kyoto Univ.}
  \textbf{35} (1999), 321--390, \href{http://arxiv.org/abs/q-alg/9603015}{{\tt
  arXiv:q-alg/9603015}}.

\bibitem{GouZhaBra93}

M.~D. Gould, R.~B. Zhang, and A.~J. Bracken, \emph{Quantum double construction
  for graded {H}opf algebras},
  \href{http://dx.doi.org/10.1017/S0004972700015197}{Bull. Aust. Math. Soc.}
  \textbf{47} (1993), 353--375.

\bibitem{HakSed94}

T.~S. Hakobyan and A.~G. Sedrakyan, \emph{Universal ${R}$ matrix of $\mathrm
  {U}_q \mathrm {sl}(n, m)$ quantum superalgebras},
  \href{http://dx.doi.org/10.1063/1.530522}{J. Math. Phys.} \textbf{35} (1994),
  2552--2559.

\bibitem{KhoTol91}

S.~M. Khoroshkin and V.~N. Tolstoy, \emph{Universal ${R}$-matrix for quantized
  (super)algebras}, \href{http://dx.doi.org/10.1007/BF02102819}{Commun. Math.
  Phys.} \textbf{141} (1991), 599--617.

\bibitem{KhoTol93a}

S.~M. Khoroshkin and V.~N. Tolstoy, \emph{The {C}artan--{W}eyl basis and the
  universal ${R}$-matrix for quantum {K}ac--{M}oody algebras and
  superalgebras}, Quantum Symmetries (H.-D. Doebner and V.~K. Dobrev, eds.),
  1993, pp.~336--351.

\bibitem{KhoTol94}

S.~Khoroshkin and V.~N. Tolstoy, \emph{Twisting of quantum (super-) algebras.
  {C}onnection of {D}rinfeld's and {C}artan-{W}eyl realizations for quantum
  affine algebras}, \href{http://arxiv.org/abs/hep-th/9404036}{{\tt
  arXiv:hep-th/9404036}}.

\bibitem{KhoTol94a}

S.~M. Khoroshlin and V.~N. Tolstoy, \emph{Twisting of quantum (super-)
  algebras}, \href{http://dx.doi.org/10.1142/9789814534314}{Generalized
  symmetries in physics} (H.-D. Doebner, V.~K. Dobrev, and A.~G. Ushveridze,
  eds.), World Scientific, 1994, pp.~42--54.

\bibitem{KhoTol92}

S.~M. Khoroshkin and V.~N. Tolstoy, \emph{The uniqueness theorem for the
  universal {$R$}-matrix}, \href{http://dx.doi.org/10.1007/BF00402899}{Lett.
  Math. Phys.} \textbf{24} (1992), 231--244.

\bibitem{ZhaGou94}

Y.-Z. Zhang and M.~D. Gould, \emph{Quantum affine algebras and universal
  ${R}$-matrix with spectral parameter},
  \href{http://dx.doi.org/10.1007/BF00750144}{Lett. Math. Phys.} \textbf{31}
  (1994), 101--110, \href{http://arxiv.org/abs/hep-th/9307007}{{\tt
  arXiv:hep-th/9307007}}.

\bibitem{BraGouZhaDel94}

A.~J. Bracken, M.~D. Gould, Y.-Z. Zhang, and G.~W. Delius, \emph{Infinite
  families of gauge-equivalent {$R$}-matrices and gradations of quantized
  affine algebras}, \href{http://dx.doi.org/10.1142/S0217979294001585}{Int. J.
  Mod. Phys. B} \textbf{8} (1994), 3679--3691,
  \href{http://arxiv.org/abs/hep-th/9310183}{{\tt arXiv:hep-th/9310183}}.

\bibitem{BraGouZha95}

A.~J. Bracken, M.~D. Gould, and Y.-Z. Zhang, \emph{Quantised affine algebras
  and parameter-dependent {$R$}-matrices},
  \href{http://dx.doi.org/10.1017/S0004972700014040}{Bull. Austral. Math. Soc.}
  \textbf{51} (1995), 177--194.

\bibitem{BooGoeKluNirRaz10}

H.~Boos, F.~G{\"o}hmann, A.~Kl{\"u}mper, Kh.~S. Nirov, and A.~V. Razumov,
  \emph{Exercises with the universal {$R$}-matrix},
  \href{http://dx.doi.org/10.1088/1751-8113/43/41/415208}{J. Phys. A: Math.
  Theor.} \textbf{43} (2010), 415208 (35pp),
  \href{http://arxiv.org/abs/1004.5342}{{\tt arXiv:1004.5342 [math-ph]}}.

\bibitem{BooGoeKluNirRaz11}

H.~Boos, F.~G{\"o}hmann, A.~Kl{\"u}mper, Kh.~S. Nirov, and A.~V. Razumov,
  \emph{On the universal ${R}$-matrix for the {I}zergin--{K}orepin model},
  \href{http://dx.doi.org/10.1088/1751-8113/44/35/355202}{J. Phys. A: Math.
  Theor.} \textbf{44} (2011), 355202 (25pp),
  \href{http://arxiv.org/abs/1104.5696}{{\tt arXiv:1104.5696 [math-ph]}}.

\bibitem{BazTsu08}

V.~V. Bazhanov and Z.~Tsuboi, \emph{Baxter's {Q}-operators for supersymmetric
  spin chains}, \href{http://dx.doi.org/10.1016/j.nuclphysb.2008.06.025}{Nucl.
  Phys. B} \textbf{805} (2008), 451--516,
  \href{http://arxiv.org/abs/0805.4274}{{\tt arXiv:0805.4274 [hep-th]}}.

\bibitem{BooGoeKluNirRaz13}

H.~Boos, F.~G{\"o}hmann, A.~Kl{\"u}mper, Kh.~S. Nirov, and A.~V. Razumov,
  \emph{Universal integrability objects},
  \href{http://dx.doi.org/10.1007/s11232-013-0002-8}{Theor. Math. Phys.}
  \textbf{174} (2013), 21--39, \href{http://arxiv.org/abs/1205.4399}{{\tt
  arXiv:1205.4399 [math-ph]}}.

\bibitem{Raz13}

A.~V. Razumov, \emph{Monodromy operators for higher rank},
  \href{http://dx.doi.org/10.1088/1751-8113/46/38/385201}{J. Phys. A: Math.
  Theor.} \textbf{46} (2013), 385201 (24pp),
  \href{http://arxiv.org/abs/1211.3590}{{\tt arXiv:1211.3590 [math.QA]}}.

\bibitem{BazHibKho02}

V.~V. Bazhanov, A.~N. Hibberd, and S.~M. Khoroshkin, \emph{Integrable structure
  of {$\mathcal W_3$} conformal field theory, quantum {B}oussinesq theory and
  boundary affine {T}oda theory},
  \href{http://dx.doi.org/10.1016/S0550-3213(01)00595-8}{Nucl. Phys. B}
  \textbf{622} (2002), 475--574,
  \href{http://arxiv.org/abs/hep-th/0105177}{{\tt arXiv:hep-th/0105177}}.

\bibitem{Koj08}

T.~Kojima, \emph{Baxter's ${Q}$-operator for the ${W}$-algebra ${W_N}$},
  \href{http://dx.doi.org/10.1088/1751-8113/41/35/355206}{J. Phys. A: Math.
  Theor} \textbf{41} (2008), 355206 (16pp),
  \href{http://arxiv.org/abs/0803.3505}{{\tt arXiv:0803.3505 [nlin.SI]}}.

\bibitem{BooGoeKluNirRaz14b}

H.~Boos, F.~G{\"o}hmann, A.~Kl\"umper, Kh.~S. Nirov, and A.~V. Razumov,
  \emph{Quantum groups and functional relations for higher rank},
  \href{http://dx.doi.org/10.1088/1751-8113/47/27/275201}{J. Phys. A: Math.
  Theor.} \textbf{47} (2014), 275201 (47pp),
  \href{http://arxiv.org/abs/1312.2484}{{\tt arXiv:1312.2484 [math-ph]}}.

\bibitem{NirRaz16a}

Kh.~S. Nirov and A.~V. Razumov, \emph{Quantum groups and functional relations
  for lower rank}, \href{http://dx.doi.org/10.1016/j.geomphys.2016.10.014}{J.
  Geom. Phys.} \textbf{112} (2017), 1--28,
  \href{http://arxiv.org/abs/1412.7342}{{\tt arXiv:1412.7342 [math-ph]}}.

\bibitem{BooJimMiwSmiTak07}

H.~Boos, M.~Jimbo, T.~Miwa, F.~Smirnov, and Y.~Takeyama, \emph{Hidden
  {G}rassmann structure in the {XXZ} model},
  \href{http://dx.doi.org/10.1007/s00220-007-0202-x}{Commun. Math. Phys.}
  \textbf{272} (2007), 263--281,
  \href{http://arxiv.org/abs/hep-th/0606280}{{\tt arXiv:hep-th/0606280}}.

\bibitem{BooJimMiwSmiTak09}

H.~Boos, M.~Jimbo, T.~Miwa, F.~Smirnov, and Y.~Takeyama, \emph{Hidden
  {G}rassmann structure in the {XXZ} model {II}: {C}reation operators},
  \href{http://dx.doi.org/10.1007/s00220-008-0617-z}{Commun. Math. Phys.}
  \textbf{286} (2009), 875--932, \href{http://arxiv.org/abs/0801.1176}{{\tt
  arXiv:0801.1176 [hep-th]}}.

\bibitem{BooJimMiwSmi10}

H.~Boos, M.~Jimbo, T.~Miwa, and F.~Smirnov, \emph{Hidden {G}rassmann structure
  in the {XXZ} model {IV}: {CFT} limit},
  \href{http://dx.doi.org/10.1007/s00220-010-1051-6}{Commun. Math. Phys.}
  \textbf{299} (2010), 825--866, \href{http://arxiv.org/abs/0911.3731}{{\tt
  arXiv:0911.3731 [hep-th]}}.

\bibitem{KluNirRaz20}

A.~Kl\"umper, Kh.~S. Nirov, and A.~V. Razumov, \emph{Reduced {qKZ} equation:
  general case}, \href{http://dx.doi.org/10.1088/1751-8121/ab3b9e}{J. Phys. A:
  Math. Gen.} \textbf{53} (2020), 015202 (35pp),
  \href{http://arxiv.org/abs/1905.06014}{{\tt arXiv:1905.06014 [math-ph]}}.

\bibitem{Raz20}

A.~V. Razumov, \emph{Reduced {qKZ} equation and genuine {qKZ} equation},
  \href{http://dx.doi.org/10.1088/1751-8121/aba91d}{J. Phys. A: Math. Theor.}
  (2020), 405204 (32pp), \href{http://arxiv.org/abs/2004.02624}{{\tt
  arXiv:2004.02624 [math-ph]}}.

\bibitem{IpZei14}

I.~C.-H. Ip and A.~M. Zeitlin, \emph{{$Q$}-operator and fusion relations for
  {$U_q(C^{(2)}(2))$}},
  \href{http://dx.doi.org/10.1007/s11005-014-0702-5}{Lett. Math. Phys.}
  \textbf{104} (2014), 1019--1043, \href{http://arxiv.org/abs/1312.4063}{{\tt
  arXiv:1312.4063 [math.QA]}}.

\bibitem{PerSch81}

J.~H.~H. Perk and C.~L. Schultz, \emph{New families of commuting transfer
  matrices in $q$-state vertex models},
  \href{http://dx.doi.org/10.1016/0375-9601(81)90994-4}{Phys. Lett. A}
  \textbf{84} (1981), 407--410.

\bibitem{ChaPre94}

V.~Chari and A.~Pressley, \emph{A guide to quantum groups}, Cambridge
  University Press, Cambridge, 1994.

\bibitem{Kac77}

V.~Kac, \emph{Lie superalgebras},
  \href{http://dx.doi.org/10.1016/0001-8708(77)90017-2}{Adv. Mat.} \textbf{26}
  (1977), 8--96.

\bibitem{Jim86a}

M.~Jimbo, \emph{A $q$-analogue of {$\mathrm U(\mathfrak{gl}(N + 1))$}, {H}ecke
  algebra, and the {Y}ang--{B}axter equation},
  \href{http://dx.doi.org/10.1007/BF00400222}{Lett. Math. Phys.} \textbf{11}
  (1986), 247--252.

\bibitem{Leu86}

J.~van~de Leur, \emph{Contragredient {L}ie superalgebras of finite growth},
  Ph.D. thesis, Utrecht University, 1986.

\bibitem{Leu89}

J.~W. van~de Leur, \emph{A classification of contragredient {L}ie superalgebra
  of finite growth}, \href{http://dx.doi.org/10.1080/00927878908823823}{Comm.
  in Algebra} \textbf{17} (1989), 1815--1841.

\bibitem{MenTes15}

C.~Meneghelli and J.~Teschner, \emph{Integrable light-cone lattice
  discretizations from the universal ${R}$-matrix},
  \href{http://arxiv.org/abs/1504.04572}{{\tt arXiv:1504.04572 [hep-th]}}.

\bibitem{NirRaz19}

Kh.~S. Nirov and A.~V. Razumov, \emph{Vertex models and spin chains in formulas
  and pictures}, \href{http://dx.doi.org/10.3842/SIGMA.2019.068}{SIGMA}
  \textbf{15} (2019), 068 (67pp), \href{http://arxiv.org/abs/1811.09401}{{\tt
  arXiv:1811.09401 [math-ph]}}.

\bibitem{Zha14}

H.~Zhang, \emph{Representations of quantum affine superalgebras},
  \href{http://dx.doi.org/10.1007/s00209-014-1330-6}{Math. Z.} \textbf{278}
  (2014), 663--703, \href{http://arxiv.org/abs/1309.5250}{{\tt arXiv:1309.5250
  [math.QA]}}.

\bibitem{Tan92}

T.~Tanisaki, \emph{{K}illing forms, {H}arish-{C}handra homomorphisms and
  universal ${R}$-matrices for quantum algebras}, Infinite Analysis
  (A.~Tsuchiya, T.~Eguchi, and M.~Jimbo, eds.), Advanced Series in Mathematical
  Physics, vol.~16, World Scientific, Singapore, 1992, pp.~941--962.

\bibitem{Usm94}

R.~A. Usmani, \emph{Inversion of a tridiagonal {J}acobi matrix},
  \href{http://dx.doi.org/10.1016/0024-3795(94)90414-6}{Linear Algebra Appl.}
  \textbf{212/213} (1994), 413--414.

\end{thebibliography}
\end{document}